\documentclass[showpacs, pra,twocolumn,preprintnumbers ,amsmath, amssymb, superscriptaddress, aps]{revtex4-2}

\usepackage[utf8]{inputenc}
\DeclareUnicodeCharacter{202F}{\,}
\usepackage{color}
\usepackage{amsmath,amssymb}
\usepackage{pifont}
\usepackage{amssymb}  
\usepackage{bbold}
\usepackage{float}
\usepackage{subfloat}
\usepackage[squaren]{SIunits}
\usepackage{xcolor}

\usepackage[caption=false]{subfig}
\usepackage{tikz}
\usepackage{makecell}
\usepackage{subfig}
\usepackage{pifont}   
\usepackage{graphicx} 
\graphicspath{{Figures/}}
\usepackage{dcolumn}  
\usepackage{bm}       
\usepackage{multirow} 
\usepackage{placeins}
\usepackage[colorlinks]{hyperref}
\usepackage{mathtools}
\usepackage{appendix}

\begin{document}
	
	\title{Strain- and potential-controlled tunneling in monolayer
    MoS$_2$}
	
	\date{\today}
	
%
	\author{Hasna Chnafa}
\affiliation{Laboratory of Theoretical Physics, Faculty of Sciences, Choua\"ib Doukkali University, PO Box 20, 24000 El Jadida, Morocco}

    	\author{Rachid El Aitouni}
\affiliation{Laboratory of Theoretical Physics, Faculty of Sciences, Choua\"ib Doukkali University, PO Box 20, 24000 El Jadida, Morocco}

\author{Clarence Cortes}
\affiliation{Vicerrector\'ia de Investigaci\'on y Postgrado, Universidad de La Serena, La Serena 1700000, Chile}  
\author{David Laroze}
\affiliation{Instituto de Alta Investigación, Universidad de Tarapacá, Casilla 7D, Arica, Chile}

	\author{Ahmed Jellal}
	\email{a.jellal@ucd.ac.ma}
	\affiliation{Laboratory of Theoretical Physics, Faculty of Sciences, Choua\"ib Doukkali University, PO Box 20, 24000 El Jadida, Morocco}

	\begin{abstract}

We present a theoretical study of spin- and valley-resolved quantum transport 
in monolayer MoS$_2$ under the combined influence of mechanical strain and an 
external scalar potential, a combination whose simultaneous 
unexplored.  Within an effective massive Dirac Hamiltonian that incorporates intrinsic spin--orbit coupling, strain induces valley-dependent momentum shifts that lift the degeneracy between the $K$ and $K'$ valleys and strongly modify the transport characteristics. The scalar potential modifies the tunneling spectrum, leading to pronounced changes in resonant transmission, Fabry--P\'erot interference, and conductance. We show that the interplay between strain and electrostatic potential enables efficient control of both valley and spin polarization of the transmitted current. In particular, we identify a dual-knob control scheme in which the barrier width governs the frequency of conductance oscillations while strain independently controls their phase and amplitude. Furthermore, we predict electrostatic spin inversion --- a sign reversal of spin polarization achievable purely by gate tuning at finite strain, requiring no geometric reconfiguration. Depending on the strain orientation, the transmission probability and conductance can be selectively suppressed or enhanced, resulting in highly tunable valley- and spin-polarized transport. These findings demonstrate that strain and potential engineering provide orthogonal and independently operable mechanisms for controlling conductance as well as spin and valley degrees of freedom in monolayer MoS$_2$, offering promising prospects for spintronic and valleytronic device applications.


	\end{abstract}
	
	\pacs{72.80.Vp, 73.23.-b, 78.67.-n\\
		{\sc Keywords:}
		Monolayer MoS$_2$, strain, scalar potential, Dirac equation,  transmission, {Klein} tunneling, conductance, polarization.}
	\maketitle

	
	
	\section{Introduction}	\label{Intro}
Two-dimensional ($2$D) materials have recently attracted considerable attention due to their exceptional electronic and optical properties, offering promising ground for applications in nanoelectronics \cite{int1,intr1,intr2}, optoelectronics \cite{int2,intr3,intr4}, and spintronics \cite{int3,Ahn2020,Feng2017}. Among these materials, monolayer molybdenum disulfide (MoS$_2$) was isolated by exfoliation, similar to graphene and $2$D boron nitride \cite{int4}. Unlike bulk MoS$_2$, which is an indirect-gap semiconductor, monolayer MoS$_2$ has a direct band gap of approximately $1.66\,\text{eV}$ \cite{int2,int6}, located at the $K$ and $K'$ points of the Brillouin zone \cite{int3,int7}, allowing low-energy carriers to be modeled as massive Dirac fermions with a moderate spin–orbit coupling \cite{int8,int3,int9}. This gives electrons and holes an additional degree of freedom related to the valley, which can be used to encode information and for further processing \cite{int11,int12,int13,aitouni}.

MoS$_2$ exhibits high resistance to deformation, allowing them to withstand significant mechanical strain before rupture \cite{int14,int15}. The controlled application of biaxial or uniaxial strain to these monolayers makes it possible to modulate the band gap, induce a semiconductor-metal transition, and finely modify their electronic and optical properties \cite{int15,int16}. Within experimental approaches, transferring MoS$_2$ onto flexible substrates such as polydimethylsiloxane (PDMS) or polyethylene terephthalate (PET) is widely used to apply mechanical strain by stretching or compressing the substrate, enabling the measurement of spectroscopic and mechanical changes under deformation \cite{int17,int18,int19}.  Another method relies on inducing localized strain using nanoscale probe techniques, such as tip-enhanced Raman spectroscopy (TERS) or atomic force microscopy (AFM) tips, which enable very precise deformations to be applied and mapped in specific regions of MoS$_2$ \cite{int20,int22}. Lastly, thermal approaches exploit the difference in thermal expansion coefficients between MoS$_2$ and the substrate: by heating or cooling the system, a controlled biaxial strain is generated, which can be observed through characteristic shifts in the Raman and photoluminescence modes \cite{int23,int24,int25}. These techniques enable well-defined and quantifiable strain levels, providing essential tools for deformation engineering in monolayer MoS$_2$.


Furthermore, theoretical studies have shown strong interest in MoS$_2$. 
First-principles calculations indicate that reducing the dimensionality induces 
a transition from an indirect gap in the bulk material to a direct gap in 
MoS$_2$, reflecting a profound change in the electronic band structure 
\cite{int2,int6,int8,Kuc2011,Cheiwchanchamnangij2012}. DFT further confirms 
the direct nature of the band gap and the importance of orbital contributions 
in the band structure of MoS$_2$ \cite{Kadantsev2012,Dong2018,Rai2020}. 
{Beyond first-principles approaches, $k\cdot p$ methods have 
	proven particularly effective in capturing the low-energy physics of monolayer 
	MoS$_2$. Three-band and two-band spinful $k\cdot p$ Hamiltonians have been 
	derived from group-theoretical considerations and nine-band frameworks 
	\cite{Beiranvand2017,Beiranvand2018}, providing rigorous microscopic foundations 
	for the effective massive Dirac description adopted in the present work. 
	Furthermore, tight-binding investigations of uniaxial and biaxial strain effects 
	on the band gap of monolayer MoS$_2$ \cite{Shahriari2019} have demonstrated 
	that strain modifies the electronic structure in a direction-dependent manner, 
	consistent with the valley-dependent momentum shifts we report here.} Other 
work has analyzed in detail the electronic structure with spin--orbit coupling. 
In particular, the lack of inversion symmetry and the strong spin--orbit 
interaction in MoS$_2$ lead to spin--valley coupling at the $K$ and $K'$ 
{points}, which is essential for spintronics and valleytronics 
physics \cite{Li2024,Chan2024}. Although several studies have addressed 
electronic transport in MoS$_2$, the combined effect of uniaxial strain and 
an external scalar potential remains largely unexplored and constitutes the 
main focus of this work.

We study the transport of electrons through a monolayer of molybdenum 
disulfide MoS$_2$ subjected to an electrostatic barrier and uniaxial 
strain{, a combination that has not been simultaneously 
	addressed in previous theoretical works on this material}. In order to 
highlight the fundamental physics, we use an effective massive Dirac 
Hamiltonian including intrinsic spin--orbit coupling, from which we evaluate 
the eigenvalues and eigenvectors. Electron transmission is determined by 
applying continuity conditions at the barrier interfaces. This allows us to 
calculate and analyze the conductance along with the spin and valley 
polarizations of the transmitted current. As a numerical result, we show 
that the transmission exhibits pronounced Fabry--P\'erot resonances due to 
quantum interference and electron confinement in the barrier region. These 
resonances are highly dependent on the {spin and valley} 
degrees of freedom. Their position and width can be {jointly} 
modulated by {strain and} electrostatic parameters. 
{In particular, we identify a dual-knob control scheme in 
	which the barrier width governs the frequency of conductance oscillations 
	while strain independently controls their phase and amplitude. Furthermore, 
	we predict electrostatic spin inversion --- a sign reversal of spin 
	polarization achievable purely by gate tuning at finite strain, requiring 
	no geometric reconfiguration.} The deformation induces valley-dependent 
momentum shifts, lifting the degeneracy between the $K$ and $K'$ valleys 
and significantly affecting electronic transport. The scalar potential 
modifies the tunneling spectrum, leading to marked variations in resonant 
transmission and conductance. We conclude that the combination of mechanical 
deformation and electrostatic potential {provides orthogonal 
	and independently operable mechanisms for controlling spin and valley degrees 
	of freedom in monolayer MoS$_2$}, offering promising prospects for 
applications in spintronic and valleytronic devices.

The paper is structured as follows.  In Sec.~\ref{sec2}, we present the physical setup of the monolayer MoS$_2$ and introduce the model Hamiltonian used to describe the system. 
Subsequently, we formulate the theoretical framework by adopting a continuous model 
{for the electronic properties of MoS$_2$}, 
from which we derive the eigenenergies and eigenspinors for the three regions of the system. 
{Sec.~\ref{sec3} is devoted to the calculation of the transmission probability by applying the appropriate boundary conditions at the interfaces $x=0$ and $x=L$. We then analyze the numerical results and discuss the influence of the relevant physical parameters, including strain and electrostatic potential.} 
Sec.~\ref{sec4} 
{is dedicated to the study of the conductance as well as the spin and valley polarizations.} 
{In Sec.~\ref{V}, we discuss the experimental feasibility and limitations of the present theoretical model.} 
In Sec.~\ref{sec5}, we conclude by summarizing our main findings.


\section{Low-energy Hamiltonian} \label{sec2}

\begin{figure}[ht]
		\centering
		\includegraphics[scale=0.4]{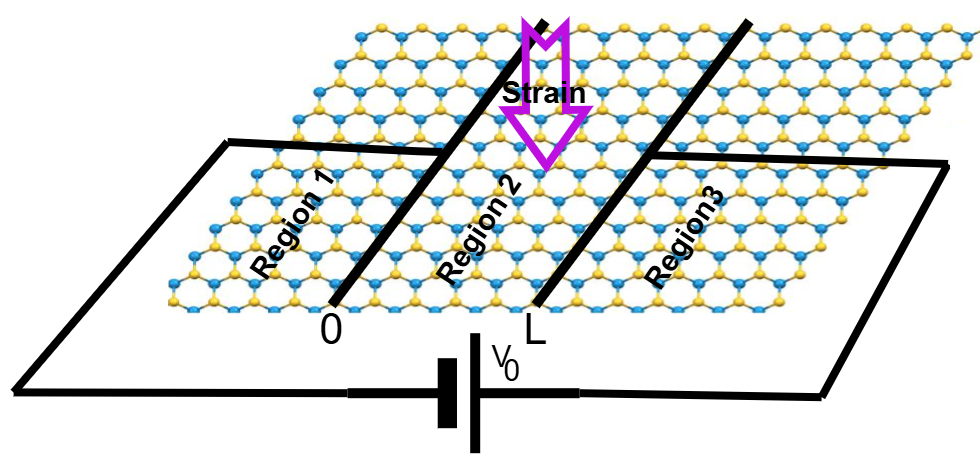}
		\caption{Schematic illustration of a molybdenum disulfide MoS$_2$  sheet under a uniaxial strain field and an external scalar electrostatic potential applied along the transport direction $x$, which divides the sheet into three distinct regions.
		}\label{str}
	\end{figure}

Our system is assumed to be translationally invariant in the $y$-direction and is divided into three regions: region I $(x < 0)$, region II $(0 \le x \le L)$, and region III $(x > L)$, as shown in Fig.~\ref{str}. The scalar potential and strain are present only in region II. In regions I and III, carriers propagate freely in an unstrained molybdenum disulfide MoS$_2$ sheet. In region II, however, they experience an electrostatic barrier and a strain-induced modification of the Dirac Hamiltonian. The strain field generates an effective gauge potential that couples with opposite signs to the two inequivalent valleys, $K$ and $K'$, leading to valley-dependent shifts in transverse momentum. 
As a result, the quasiparticle spectrum and transmission probabilities through the central region become strongly valley sensitive. By matching the wave functions at the interfaces $(x=0)$ and $(x=L)$, one can determine the reflection and transmission amplitudes and subsequently evaluate the valley-resolved conductance. This framework allows us to investigate the combined influence of scalar electrostatic gating and strain-induced gauge fields on electronic transport and the emergence of valley polarization in monolayer MoS$_2$.


Near the two inequivalent valleys \(K\) and \(K'\) (\(\tau=\pm1\)), the low-energy effective Hamiltonian of strained monolayer MoS$_2$ writes as
\begin{align}\label{ham1}
	H_\tau =& \hbar v_F \left( \tau \sigma_x k_x + \sigma_y k_y \right)
	+ \frac{\Delta}{2}\sigma_z
	- \lambda \tau s_z \frac{\sigma_z - \mathbb{I}_2}{2}\notag\\
	&+ V(x)\mathbb{I}_2
	+ H_{\text{strain}},
\end{align}
where  \( v_F = \frac{a t}{\hbar} \approx 0.53 \times 10^{6}\,\text{m/s} \) is the Fermi velocity, 
\(\Delta \simeq 1.66~\text{eV}\) is the intrinsic band gap,
 \(\lambda\simeq 75~\text{meV}\) denotes the spin--orbit coupling strength,
\(s_z=\pm1\) labels the electron spin,
\(\sigma_i\) $(i=1,2,3)$ are Pauli matrices acting in the pseudospin space, $\mathbb{I}_2$ is the matrix unit, and
the electrostatic potential $V(x)$ is modeled as a rectangular barrier
\begin{align}
	V(x) =
	\begin{cases}V_0, & 0 \le x \le L,\\
		0, & \text{otherwise}.	
	\end{cases}
\end{align}
Mechanical strain in a monolayer MoS$_2$ modifies the electronic spectrum through the emergence of an effective gauge field that couples to the Dirac fermions in a valley-dependent manner. Then, $H_{\text{strain}}$ can be expressed as
\begin{align}
	H_{\text{strain}} &= \hbar v_F \left( \tau \sigma_x A_x + \sigma_y A_y \right),
\end{align}
and
the components of strain-induced vector potential are given by
\begin{align}
	A_x = \beta \left( u_{xx} - u_{yy} \right), 
	\quad
	A_y = -2 \beta u_{xy},
\end{align}
where $u_{ij}$ denotes the strain tensor and $\beta\approx2.4$  is the material-dependent Gr\"uneisen parameter \cite{1}. 
In the present work, we focus on uniaxial strain applied along the transport direction $x$, such that $u_{xy}=0$, and only the $A_x$ component survives.
For a uniform uniaxial strain profile, we have
\begin{align}
	u_{xx} = \varepsilon, 
	\quad
	u_{yy} = -\nu \varepsilon,
\end{align}
with $\varepsilon$ is the strain amplitude and $\nu=0.25$  is the Poisson ratio \cite{2}.  
{The strain values considered here, up to $\varepsilon = 0.6$, 
	are consistent with the large elastic deformation tolerance of monolayer 
	MoS$_2$~\cite{int14,int15,int16,int17,int18,int19}, with the most physically 
	significant and experimentally relevant results obtained for 
	$\varepsilon \leq 0.3$.}
Then, the gauge field reduces to a constant
\begin{align}
	A_x = \beta \varepsilon (1+\nu).
\end{align}
This effective vector potential is assumed to be nonzero only within the barrier region ($0 \le x \le L$), while it vanishes in the source and drain regions. Physically, the gauge field $A_x$ breaks the valley degeneracy of the transmission spectrum without violating time-reversal symmetry, and provides an efficient mechanism for strain-controlled valley filtering in MoS$_2$ nanostructures.
%

To determine the solutions of the energy spectrum, we write 
the Hamiltonian \eqref{ham1}, in matrix form, as
\begin{align}
    H = \begin{pmatrix} V_{0} + \frac{\Delta}{2} &  v_{F}\Big[\tau(p_{x}+\hbar A_{x}) - ip_{y}\Big] \\  v_{F}\Big[\tau(p_{x}+\hbar A_{x}) + ip_{y}\Big] & V_{0} - \frac{\Delta}{2} + \lambda \tau s_{z} \end{pmatrix}, 
\end{align}
which acts on the spinor $\psi_{\tau s_z}(x,y)$ associated with the energy $E$ as 
	$H \psi_{\tau s_z}(x,y) = E \psi_{\tau s_z}(x,y).$
Due to translational invariance along \(y\)-direction, we have
\begin{align}
	\psi_{\tau s_z}(x,y) = e^{ik_y y}\phi_{\tau s_z}(x),
\end{align}
where $
	\phi_{\tau s_z} = 
	\begin{pmatrix}
		\psi_{A},
		\psi_B
	\end{pmatrix}
	^T$.
In region II $\, (0 \le x \le L)$, we get the two coupled equations
%
\begin{widetext}
\begin{align}
  &  \left(E - V_{0} - \frac{\Delta}{2}\right)\psi_{A}(x) = \hbar v_{F}\Big[\tau\big(-i\,\partial_{x} + A_{x}\big) - ik_{y}\Big]\psi_{B}(x),\\
 & \left(E - V_{0} + \frac{\Delta_{\tau s_z}}{2}\right)\psi_{B}(x) = \hbar v_{F}\Big[\tau\big(-i\partial_{x} + A_{x}\big) + ik_{y}\,\Big]\psi_{A}(x),
\end{align} 
\end{widetext}
where we have set $\Delta_{\tau s_z} = \Delta - 2\lambda \tau s_z$.
We can eliminate one component to obtain a second-order differential equation for  $\psi_A(x)$
\begin{align}
      \left(\partial_{x}^{2} + 2iA_{x}\partial_{x} +\kappa \,\right) \psi_{A}(x) = 0,
\end{align}
with the parameter  
\begin{align}\label{eq12}
\kappa=\frac{\left(E - V_{0} + \dfrac{\Delta_{\tau s_z}}{2}\right)\left(E - V_{0} - \dfrac{\Delta}{2}\right)}{(\hbar v_{F})^{2}}-(A_{x}^{2} + k_{y}^{2}).
\end{align}
We show the eigenspinor can be written as
\begin{align}
   \psi^{\text {II}}_{\tau s_z}(x,y) =& e^{-iA_{x}x}\left[a_{1}\begin{pmatrix}
      1\\\Gamma
 \end{pmatrix}e^{i q_x x} + a_{2}\begin{pmatrix}
      1\\-\Gamma^*
  \end{pmatrix}e^{-i q_x x}\right]\notag\\
  &e^{ik_yy},
\end{align}
where we have defined
\begin{align}
    &\Gamma=\frac{\hbar v_{F}\left[\tau\left(q_x +  A_x\right)+ i k_{y}\right]}{E - V_{0} + \dfrac{\Delta_{\tau s_z}}{2}}=e^{i\phi},\\
    &\phi = \arctan\!\left(\frac{k_y}{\tau\left(q_x + A_x\right)}\right).
\end{align}
The corresponding energy can be expressed as
	\begin{align}\label{eq3}
		E= V_0 + s' \hbar v_{F} \sqrt{(q_x + A_x)^2 + k_y^2 + \left( \frac{\Delta_{\tau s_z}}{2\hbar v_F} \right)^2 },
	\end{align}
    with $s' = \mathrm{sign}(E - V_0)$.  As a result, we obtain the longitudinal component 
\begin{align}
    &q_x=\tau\sqrt{\kappa}.
    \end{align}
 %
In region {\text {I}}  (\( x < 0 \)), there is no strain and no potential. Then, the eigenvalue equations decouple for the two spinors and take the following form
\begin{align}
    \left(\partial_{x}^{2} + \kappa'  \right)\psi_{A}(x) = 0\label{EqSans},
\end{align} 
and we have   	\begin{align}\kappa'=\frac{\left(E + \dfrac{\Delta_{\tau s_z}}{2}\right)\left(E - \dfrac{\Delta}{2}\right)}{(\hbar v_{F})^{2}}- k_{y}^{2}.
\end{align}
The solution is
\begin{align}
    \psi^{\text {I}}_{\tau s_z}(x,y) = \left[\begin{pmatrix}
        1\\\gamma
    \end{pmatrix}e^{i k_x x} + r\begin{pmatrix}
        1\\-\gamma^*
    \end{pmatrix}e^{-i k_x x}\right]e^{ik_yy},
\end{align}
where $r$ is the reflection coefficient, $\gamma$ and $\theta$ are
\begin{align}
   	&\gamma=\frac{\hbar v_{F} \left[\tau k_x + i k_{y}\right]}{E + \dfrac{\Delta_{\tau s_z}}{2}}=e^{i\theta},\\
   	&\theta = \arctan\!\left(\frac{k_y}{\tau k_x }\right).
\end{align}
The associated energy is \begin{align}
		E=  s\hbar v_{F} \sqrt{k^2 + \left( \frac{\Delta_{\tau s_z}}{2\hbar v_F} \right)^2 },
\end{align}
with  $s=\text{sign}(E)$, and $k=\sqrt{k^2_x + k^2_y}$. We can derive the wave  vector component $k_x$ as
\begin{align}
	&k_x=\tau\sqrt{\kappa'}.
\end{align} 
Finally, in the region {\text {III}} (\( x > L \)), the spinor can be  written as 
\begin{align}
    \psi^{\text {III}}_{\tau s_z}(x,y)=t\begin{pmatrix}
        1\\ \gamma
    \end{pmatrix}e^{ik_xx}e^{ik_yy},
\end{align}
where $t$ is the transmission coefficient.

\begin{figure}[ht]
\subfloat[]{\includegraphics[scale=0.28]{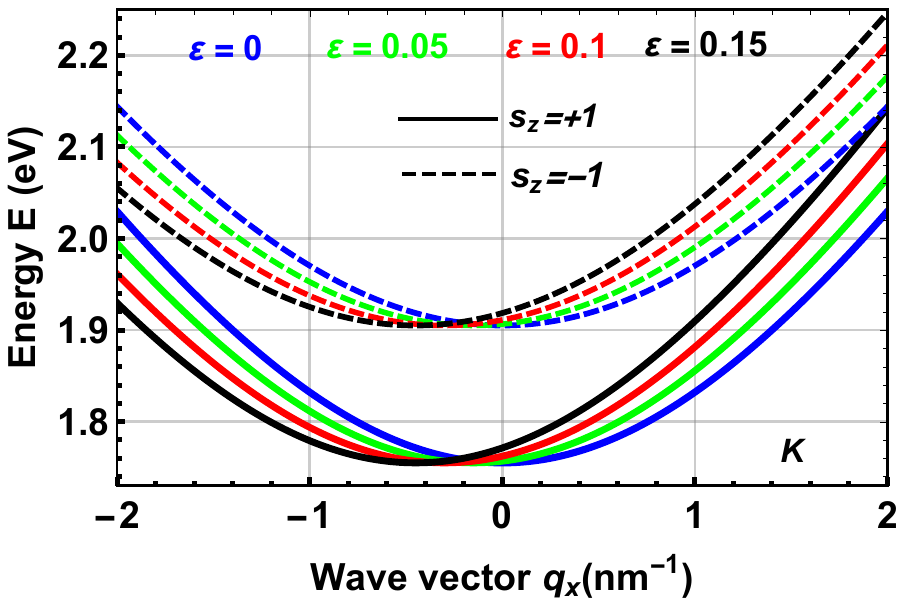}}
\subfloat[]{\includegraphics[scale=0.28]{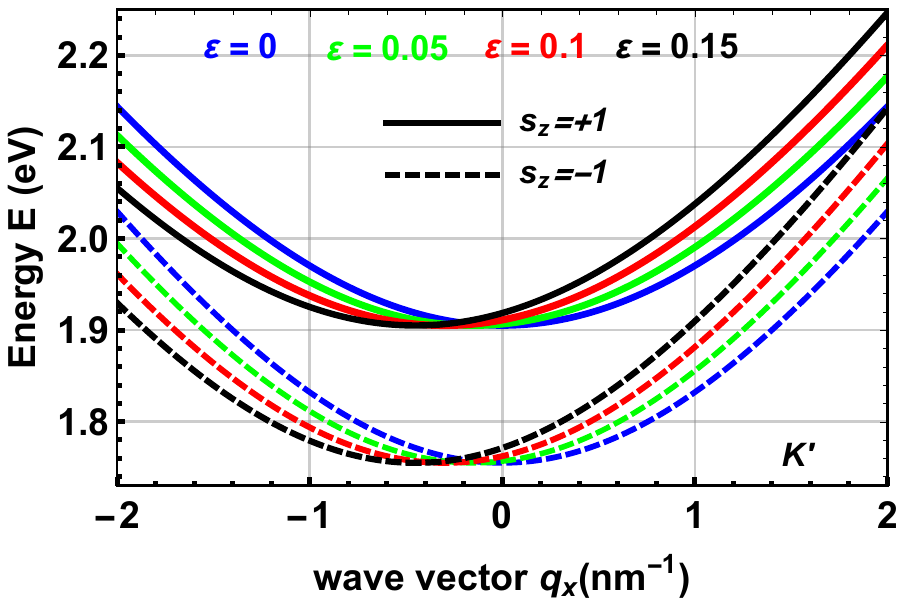}}
\caption{The energy $E$ (\ref{eq3}) of region II versus the wave vector $q_x$ for barrier height $V_0 = 1$ eV and different strain values  $\varepsilon = 0, 0.05, 0.1, 0.15$. The $K$ valley is displayed in
 (a) while $K'$ in (b).}
\label{z1}
\end{figure}


Figure~\ref{z1} displays the energy  $E$ in region II versus the wave vector $q_x$  for a barrier height $V_0 = 1$ eV and  different strain values  $\varepsilon$, with (a):   valley $K$ and (b): while  valley $K'$. In both cases, the bands exhibit a parabolic dispersion characteristic of massive carriers in MoS$_2$, with a clear separation between the spin states $s_z = +1$ and $s_z = -1$, due to strong spin-orbit coupling \cite{int3,3,Cheng2015,5}. 
This separation depends on the valley: the order of spin levels is reversed between $K$ and $K'$, which is a direct signature of spin–valley locking in transition metal dichalcogenides. The increase in deformation $\varepsilon$ causes an overall shift of the bands towards higher energies and a slight change in their curvature, reflecting a change in the effective mass of the carriers.  This joint dependence on spin, valley, and strain breaks the degeneracy between $K$ and $K'$ and favors significant valley polarization, while the contributions of the two spins can partially compensate each other, leading to weaker spin polarization.  These results demonstrate that deformation and barrier potential are powerful external parameters for engineering spin–valley transport in MoS$_2$, paving the way for potential applications in valleytronics and spintronics in two-dimensional materials.

\section{Transmission} \label{sec3}

To determine the coefficients \(r\), \(a_1\), \(a_2\), and \(t\), we apply  the boundary conditions to the eigenspinors in ensuring their continuities at the interfaces \(x = 0\) and \(x = L\).  Physically, this continuity ensures that the eigespinor does not exhibit unphysical jumps when the particle crosses from one region to another. Once these matching conditions are enforced, the scattering problem is fully constrained. The amplitudes of the reflected, transmitted, and intermediate states are no longer free parameters, but rather, they are fixed by the system's structure and length.
 This approach enables us to obtain a complete and self-consistent description of the transport process, allowing us to compute observable quantities—such as reflection and transmission probabilities—in a transparent and physically intuitive way. Then, we write
\begin{align}
	&\psi^{\text {I}}_{\tau s_z}(0,y) = \psi^{\text {II}}_{\tau s_z}(0,y),\\
	&\psi^{\text {II}}_{\tau s_z}(L,y) = \psi^{\text {III}}_{\tau s_z}(L,y),
\end{align}
from which we get
\begin{align}
    &1+r_{\tau s_z}=a_1+a_2,\\
    &\gamma-r_{\tau s_z}\,\gamma^*=\Gamma\, a_1- \Gamma^*\, a_2,\\
    &e^{-i A_xL}(a_1e^{iq_xL}+a_2e^{-iq_xL})=t_{\tau s_z}\,e^{ik_xL},\\
    &e^{-i A_xL}(a_1\Gamma\,e^{iq_xL}-a_2\Gamma^*\,e^{-iq_xL})=t_{\tau s_z}\gamma\,e^{ik_xL}.
\end{align}
These can be solved to obtain the reflection and transmission coefficients
\begin{widetext}
\begin{align}
	r_{\tau   s_z} &=\dfrac{4 \, e^{i(q_x L + 2\theta + \phi)} \, \sin(q_x L) \, (\sin\theta - \sin\phi)}
	{e^{2 i \phi} (e^{2 i q_x L} + e^{2i\theta}) + 1 + e^{2i(q_x L + \theta)} - 2 e^{i(\theta+\phi)}(-1 + e^{2 i q_x L}) }\\
	t_{\tau   s_z}& = \dfrac{e^{-i L (A_x+ k_x) } \cos\theta \cos\phi}{\cos(q_x L) \cos\theta \cos\phi - i \, \sin(q_x L) (1 - \sin\theta \sin\phi)} \, 
\end{align}\end{widetext}
The transmission probability can be calculated using the ratio of the transmitted and incident current densities. Performing this calculation in an explicit manner reveals that all velocity factors are negated, thereby demonstrating that the transmission is contingent exclusively on the modulus squared of the transmission amplitude. Then, we obtain
\begin{align}\label{TTPP}
	T_{\tau s_z}(E,k_y, V_0, L, \varepsilon) =
	\left| t_{\tau   s_z} \right|^2,
\end{align}
and after straightforward algebra, we derive
\begin{widetext}
   \begin{align}\label{eq35}
T_{\tau s_z}(E,k_y, V_0, L, \varepsilon)=\frac{\cos^{2}\theta\,\cos^{2}\phi}
{\cos^{2}\theta\,\cos^{2}\phi\,\cos^{2}(q_x L)
+ \sin^{2}(q_x L)\left(1-\sin\theta\,\sin\phi\right)^{2}}.
\end{align} 
\end{widetext}
The following is a comprehensive numerical investigation of how mechanical strain, barrier potential, incident energy, and barrier width affect the transport properties of the system. Specifically, we analyze how these parameters influence electronic transmission, conductance, and spin and valley polarization. This analysis sheds light on the combined role of strain engineering and barrier parameters in controlling spin- and valley-dependent transport in monolayer MoS$_2$.

\begin{figure}[ht]
\subfloat[]{\includegraphics[scale=0.28]{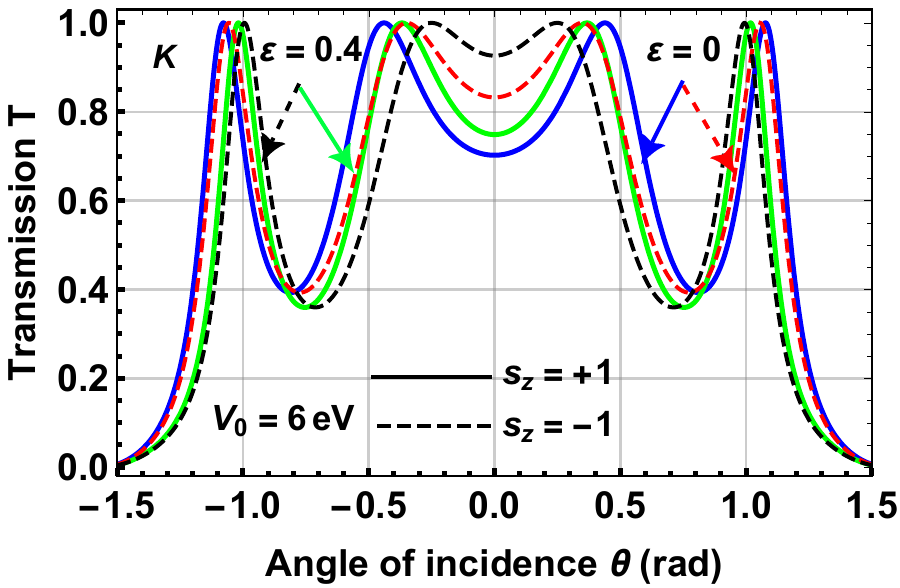}\label{f3a}}
\subfloat[]{\includegraphics[scale=0.28]{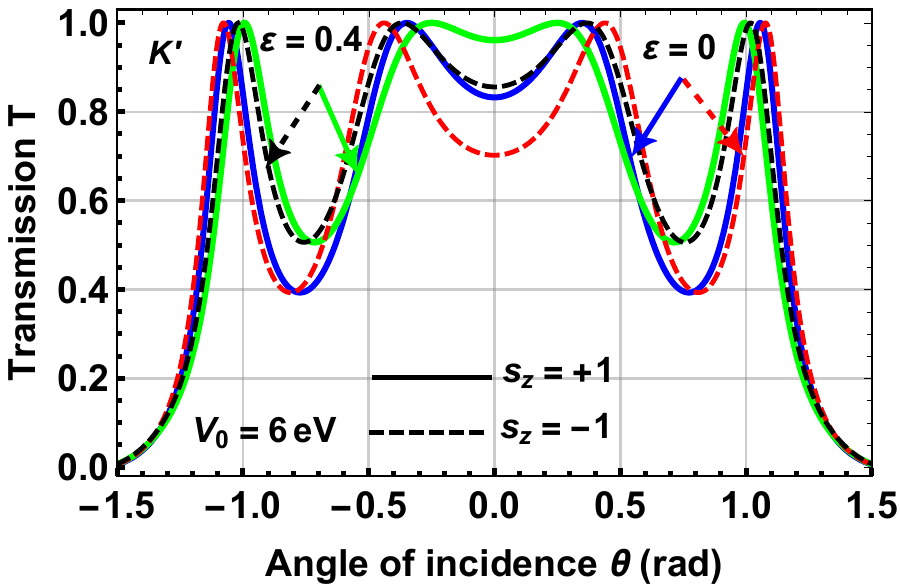}\label{f3b}}\\
\subfloat[]{\includegraphics[scale=0.28]{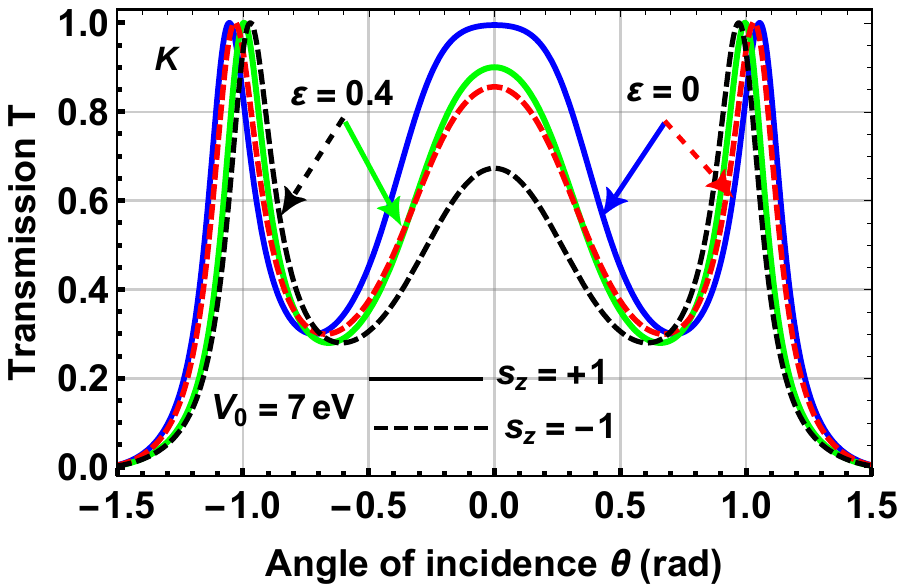}\label{f3c}}
\subfloat[]{\includegraphics[scale=0.28]{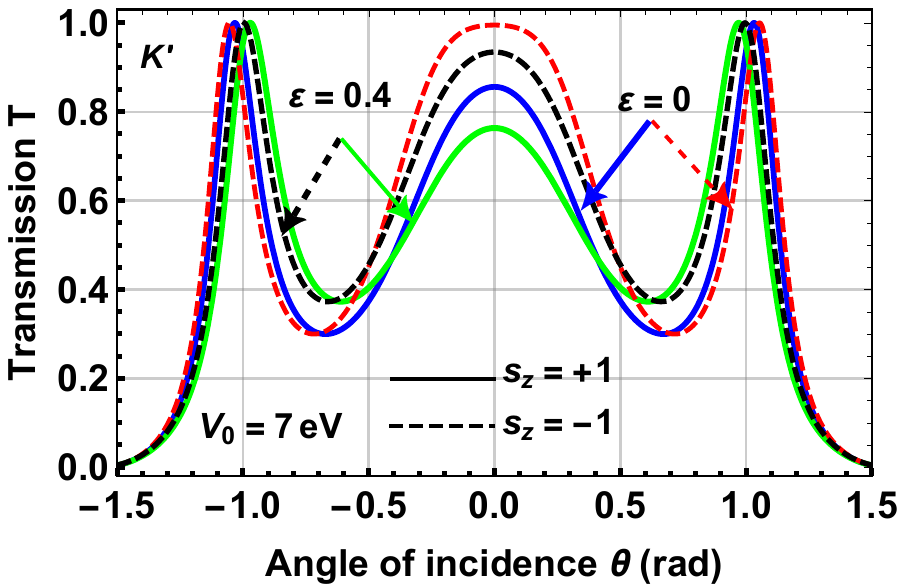}\label{f3d}}
\caption{Transmission probability  versus the angle of incidence $\theta$ in $K$ and $K'$ valleys for $E=2$ eV, $L=4$ \text{nm}, $\varepsilon=0,0.4$  and two barrier height values \text{{(a)/(b)}}: $V_0=6$  eV, \text{{(c)/(d)}}: $V_0=7$  eV. }\label{za}
\end{figure}

Figure~\ref{za} illustrates the transmission probability versus the angle 
of incidence $\theta$ for electrons crossing a barrier in a monolayer of 
MoS$_2$, taking into account the effects of spin, valley ($K, K'$), 
deformation $\varepsilon$, and barrier height $V_0$. Figs. \ref{f3a} and 
\ref{f3b}, corresponding respectively to valleys $K$ and $K'$ for $V_0 = 6$ 
eV, show quasi-unit transmission peaks at oblique angles, associated with 
quantum resonances in the barrier, while a reduction in transmission appears 
near normal incidence \cite{4,Cheng2015,5}. {This suppression 
	at normal incidence is a direct consequence of the finite band gap $\Delta$ 
	in the massive Dirac spectrum of MoS$_2$, which prevents perfect Klein 
	tunneling and forces resonant transmission to occur only at oblique angles 
	where the phase-matching condition $q_x L = n\pi$ is satisfied for integer 
	$n$.} The introduction of strain modifies the position and amplitude of these 
peaks {by shifting the longitudinal wave vector $q_x$ inside 
	the barrier through the strain-induced gauge field $A_x = \beta\varepsilon
	(1+\nu)$, which enters the dispersion relation asymmetrically for the two 
	valleys} and enhances the difference between the spin contributions $s_z = +1$ 
and $s_z = -1$, with a clear reversal of spin behavior between the $K$ and 
$K'$ valleys{. This reversal is a direct manifestation of 
	spin--valley locking: the spin--orbit coupling term $-\lambda\tau s_z$ in the 
	Hamiltonian carries opposite signs for $\tau = +1$ ($K$) and $\tau = -1$ 
	($K'$), so that the spin subband that lies higher in energy at $K$ lies lower 
	at $K'$, producing the observed mirror symmetry between the two valleys.} 
When the barrier height is increased to $V_0 = 7$ eV, as shown in 
Figs. \ref{f3c} and \ref{f3d}, the resonances become narrower and the angular 
dependence of the transmission is enhanced, amplifying the spin and valley 
asymmetry, particularly under strain. {Physically, raising 
	$V_0$ increases the mismatch between the wave vectors in the barrier and 
	incident regions, $q_x$ and $k_x$ respectively, which reduces the number of 
	angles satisfying the resonance condition and narrows each resonance peak in 
	$\theta$-space. The effective barrier seen by each spin--valley channel 
	$(\tau, s_z)$ differs because $q_x$ depends on $\Delta_{\tau s_z} = \Delta - 
	2\lambda\tau s_z$, so that the four channels acquire distinct resonance 
	angles, producing the amplified spin and valley asymmetry observed under 
	strain.} This behavior can be explained by stronger confinement of charge 
carriers inside the barrier, which increases the wave vector shift between 
the incident and barrier regions and limits transmission to very specific 
resonance conditions. As a result, the angular dependence of transmission is 
enhanced, making electron transport highly sensitive to $\theta$. Moreover, 
the increased barrier height amplifies quantum interference effects, so that 
only electrons satisfying strict phase-matching conditions 
{$\cos(q_x L)\cos\theta\cos\phi - i\sin(q_x L)(1 - \sin\theta
	\sin\phi) = 0$} can tunnel through the barrier{, confirming 
	that transmission selectivity arises from the interplay between the 
	spin--valley-dependent dispersion and the electrostatic confinement length 
	$L$.}

\begin{figure}[ht]
\subfloat[]{\includegraphics[scale=0.28]{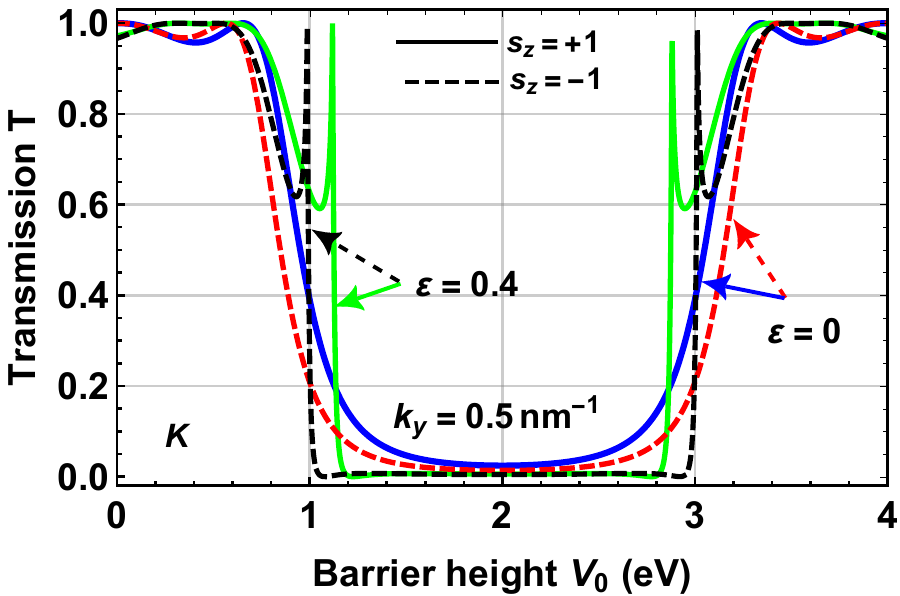}\label{fig3a}}
\subfloat[]{\includegraphics[scale=0.28]{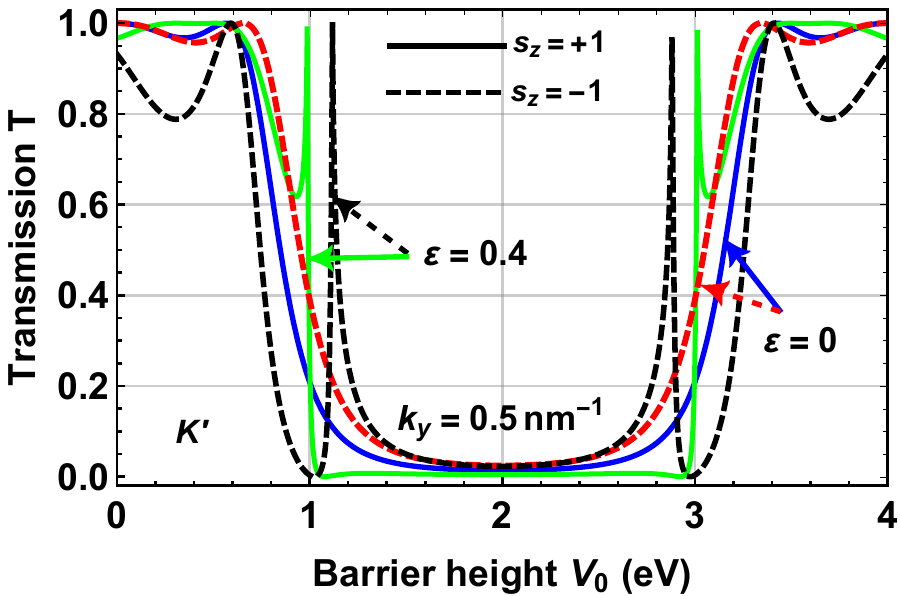}\label{fig3b}}\\
\subfloat[]{\includegraphics[scale=0.28]{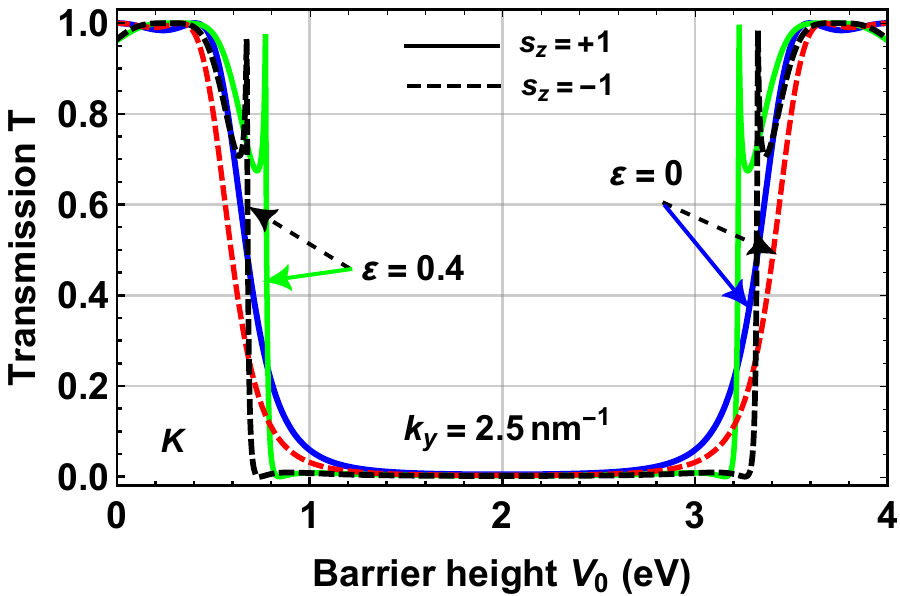}\label{fig3c}}
\subfloat[]{\includegraphics[scale=0.28]{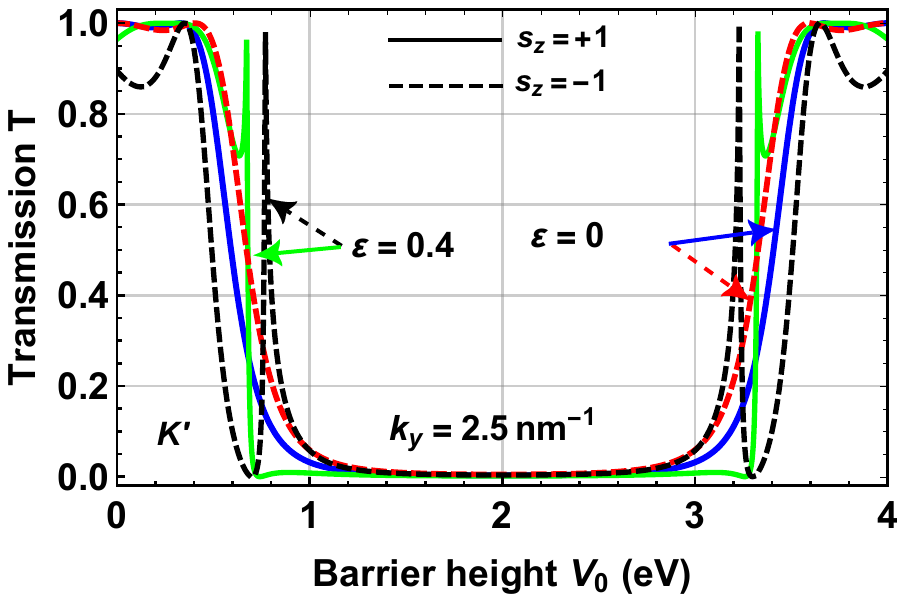}\label{fig3d}}
\caption{Transmission probability versus the barrier height $V_0$ in $K$ and $K'$ valleys for $E=2$ eV, $L=1$ \text{nm}, $\varepsilon=0,0.4$ and two wave vector values \text{{(a)/(b)}}: $k_y=0.5$ \text{nm$^{-1}$}, \text{{(c)/(d)}}: $k_y=2.5$ \text{nm$^{-1}$}}.\label{zc}
\end{figure}

To demonstrate the effects of the deformation $\varepsilon$, the barrier 
height $V_0$, and the transverse wave vector $k_y$ on electron transport, 
Fig.~\ref{zc} shows the transmission probability as a function of the barrier 
height $V_0$ in the $K$ and $K'$ valleys for an incident energy $E = 2$~eV, 
a barrier width $L = 1$~nm, two values of deformation $\varepsilon = 0$ and 
$0.4$, two wave vectors $k_y$, and different spin states. For the strainless 
case ($\varepsilon = 0$), as shown in Figs.~\ref{fig3a} and \ref{fig3b}, the 
transmission remains nearly symmetric with respect to the two valleys and spin 
orientations. A strong suppression of the transmission is observed around 
$V_0 \simeq E$, which corresponds to an evanescent propagation regime inside 
the barrier{: when $V_0 \approx E$, the longitudinal wave 
	vector inside the barrier $q_x = \tau\sqrt{\kappa}$ becomes purely imaginary, 
	as $\kappa < 0$ in Eq.~(\ref{eq12}), so that the wave function decays 
	exponentially across the barrier and transmission is strongly suppressed.} 
This behavior is consistent with previous theoretical results reported 
in~\cite{Cheng2015,5,Yuan2018}. When deformation is applied ($\varepsilon = 
0.4$), a pronounced asymmetry emerges between the $K$ and $K'$ valleys. 
{This asymmetry originates directly from the valley-dependent 
	gauge field $A_x = \beta\varepsilon(1+\nu)$, which enters the barrier 
	dispersion relation as $q_x + A_x$ for valley $K$ ($\tau = +1$) and $q_x - 
	A_x$ for valley $K'$ ($\tau = -1$), effectively shifting the wave vector in 
	opposite directions for the two valleys and breaking their degeneracy.} In 
the $K$ valley, deformation facilitates the opening of transmission channels, 
resulting in a significant enhancement of the transmission probability for 
certain ranges of the barrier height $V_0$, particularly for a low wave 
vector $k_y = 0.5~\mathrm{nm}^{-1}$. {Physically, the 
	positive shift $q_x \rightarrow q_x + A_x$ in the $K$ valley moves the 
	system away from the evanescent regime, restoring real-valued $q_x$ and 
	reopening propagating channels for specific spin orientations determined by 
	$\Delta_{\tau s_z} = \Delta - 2\lambda\tau s_z$.} This indicates that 
mechanical strain effectively increases the barrier transparency for specific 
spin orientations ($s_z = \pm 1$) and transverse momenta, offering a 
mechanism for selective control of electron transport. In contrast, in the 
$K'$ valley, {the opposite shift $q_x \rightarrow q_x - A_x$ 
	pushes the system deeper into the evanescent regime, reducing $\kappa$ and 
	further suppressing transmission,} highlighting a clear strain-induced valley 
asymmetry. Moreover, the transmission resonances exhibit a noticeable 
broadening. {This broadening reflects a strain-induced 
	modification of the resonance condition $q_x L = n\pi$: since $A_x$ shifts 
	$q_x$ differently for each valley, the resonance peaks in $V_0$-space are 
	displaced and their widths, which scale as $\Delta V_0 \propto \hbar v_F / L$, 
	are altered through the effective change in the barrier traversal wave 
	vector.} Figs.~\ref{fig3c} and \ref{fig3d} correspond to a larger wave vector 
$k_y = 2.5~\mathrm{nm}^{-1}$. In this regime, the transmission probability 
is strongly suppressed over a wider range of barrier heights $V_0${, 
	because the larger transverse momentum increases the minimum energy required 
	for propagation inside the barrier, effectively widening the evanescent window 
	in $V_0$-space through the relation $\kappa = [(E-V_0+\Delta_{\tau s_z}/2)
	(E-V_0-\Delta/2)/(\hbar v_F)^2] - (A_x^2 + k_y^2)$, where the $k_y^2$ term 
	now dominates.} Consequently, the transmission gap becomes significantly 
broader compared to the low-$k_y$ case. Furthermore, the effect of deformation 
($\varepsilon = 0.4$) is considerably amplified at large $k_y${, 
	since the valley-splitting $2A_x$ induced by strain becomes comparable to the 
	renormalized longitudinal wave vector $\sqrt{\kappa + k_y^2}$, making the 
	relative contrast between $K$ and $K'$ channels much larger and reinforcing 
	the valley filtering efficiency.} In particular, the transmission is strongly 
enhanced in one valley while being significantly reduced in the other, 
demonstrating an efficient strain-induced valley filtering effect.

\begin{figure}[ht]
\subfloat[]{\includegraphics[scale=0.28]{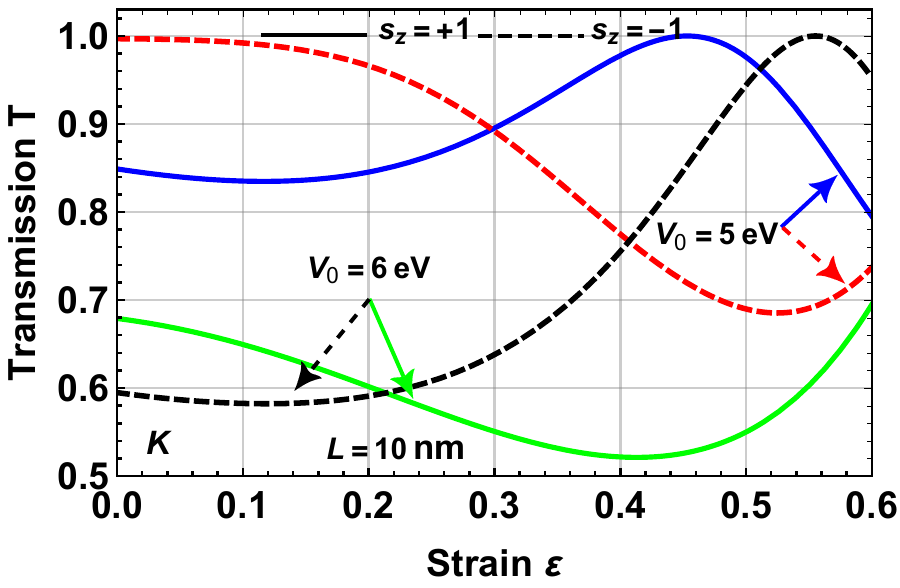}\label{f5a}}
\subfloat[]{\includegraphics[scale=0.28]{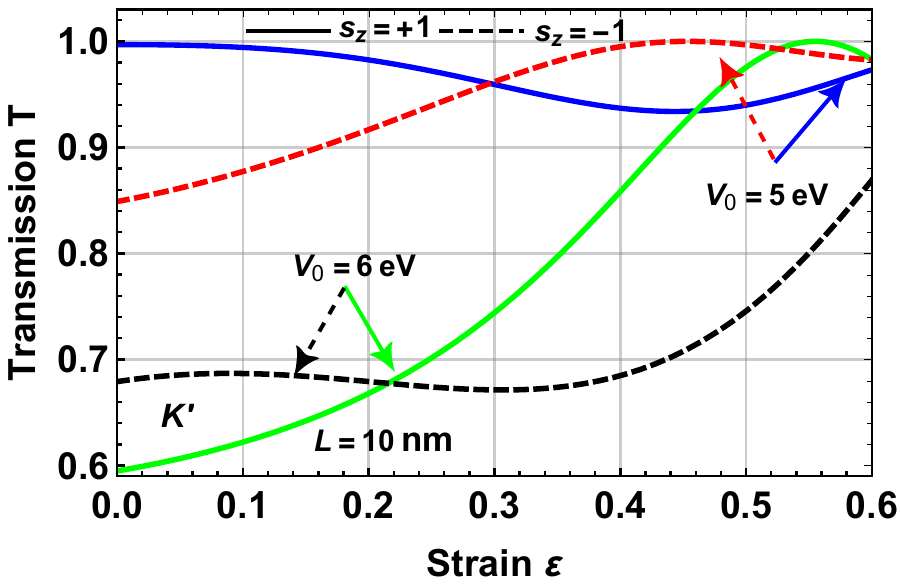}\label{f5b}}\\\subfloat[]{\includegraphics[scale=0.28]{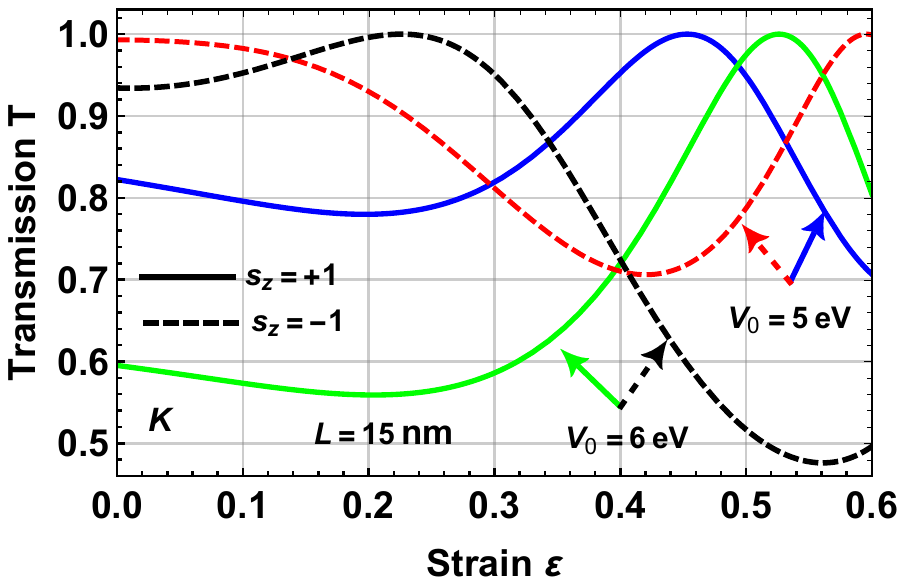}\label{f5c}}
\subfloat[]{\includegraphics[scale=0.28]{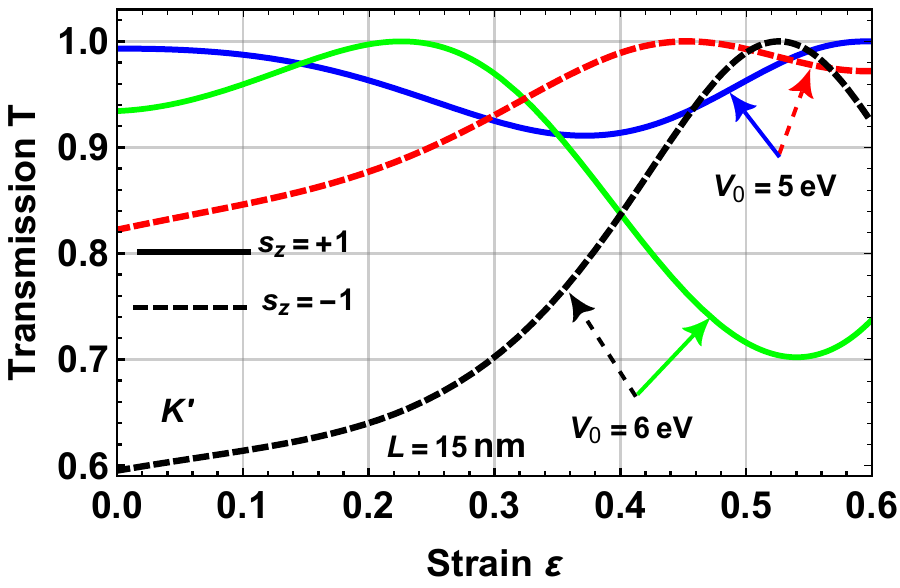}\label{f5d}}
\caption{Transmission probability versus the strain $\varepsilon$ in $K$ and $K'$ valleys for $E=2$ eV,  $k_y=0.5$ \text{nm$^{-1}$}, $V_0=5$ \text{eV}, $6$ \text{eV} and two barrier width values  \text{{(a)/(b)}}: $L=10$ \text{nm}, \text{{(c)/(d)}}: $L=15$ \text{nm}.}\label{z}
\end{figure}

To show the combined effect of mechanical deformation, barrier width, and 
barrier height, we plot Fig.~\ref{z}, which represents the evolution of the 
electronic transmission probability versus the strain $\varepsilon$ in the 
$K$ and $K'$ valleys for different values of the barrier width $L$ and the 
barrier height $V_0$. Figs.~\ref{f5a} and~\ref{f5b} correspond to a barrier 
with a smaller width, while Figs.~\ref{f5c} and~\ref{f5d} illustrate the 
case of a wider barrier. Indeed, we observe that deformation significantly 
modifies transmission, introducing a strong dependence on spin and valley. 
{This dependence arises because strain enters the Hamiltonian 
	as a valley-dependent gauge field $A_x = \beta\varepsilon(1+\nu)$, which 
	shifts the longitudinal wave vector inside the barrier as $q_x \rightarrow 
	q_x + \tau A_x$, where $\tau = \pm 1$ for the $K$ and $K'$ valleys 
	respectively. As $\varepsilon$ increases continuously, this shift drives the 
	system through successive Fabry--P\'erot resonance conditions $q_x L = n\pi$, 
	producing the oscillatory transmission pattern observed as a function of 
	$\varepsilon$.} Physically, the deformation acts on the band structure and 
spin--valley coupling, thereby modifying the resonance and tunneling conditions 
across the barrier{: the effective energy available for 
	longitudinal propagation in each spin--valley channel $(\tau, s_z)$ is 
	governed by $\Delta_{\tau s_z} = \Delta - 2\lambda\tau s_z$, so that the four 
	channels $(\tau = \pm 1, s_z = \pm 1)$ satisfy the resonance condition at 
	distinct values of $\varepsilon$, generating the spin- and valley-resolved 
	oscillation pattern.} Further, raising the barrier height generally reduces 
the overall electron transmission{, since a higher $V_0$ 
	increases the wave vector mismatch between the incident region ($k_x$) and 
	the barrier region ($q_x$), reducing the prefactor $\cos\theta\cos\phi$ in 
	the transmission formula (Eq.~(\ref{eq35})) and shifting the evanescent 
	window to higher energies,} although local increases in transmission can 
occur due to resonant states. At the same time, it makes the transmission 
more selective with respect to spin and valley, {because 
	raising $V_0$ narrows the set of $(\varepsilon, k_y)$ pairs for which 
	$q_x L = n\pi$ is satisfied, so that only channels with the correct 
	combination of spin--orbit energy $\Delta_{\tau s_z}$ and strain-shifted 
	wave vector $q_x + \tau A_x$ can fulfill the resonance condition and 
	contribute to transmission.} Additionally, increasing the barrier width 
enhances the tunnel regime and amplifies quantum interference effects, 
causing pronounced oscillations characteristic of Fabry--P\'erot interference 
\cite{6,7}, which result from multiple reflections of the electron waves 
inside the barrier. {Quantitatively, a wider barrier $L$ 
	compresses the resonance spacing in $\varepsilon$-space: since the resonance 
	condition requires $q_x(\varepsilon) L = n\pi$, a larger $L$ means that a 
	smaller increment in $\varepsilon$ — and hence in $A_x$ — is sufficient to 
	advance from one resonance to the next, increasing the oscillation frequency 
	and producing the denser fringe pattern observed in Figs.~\ref{f5c} 
	and~\ref{f5d} relative to Figs.~\ref{f5a} and~\ref{f5b}.} These effects 
create a greater asymmetry between the $K$ and $K'$ valleys as well as more 
effective spin selectivity for certain values of $\varepsilon${, 
	confirming that $L$ and $V_0$ act as independent control parameters: $L$ 
	governs the density of resonances in $\varepsilon$-space while $V_0$ controls 
	their contrast and spin--valley selectivity.}

\section{Conductance and Polarization} \label{sec4}

The conductance is a fundamental physical quantity for characterizing electronic transport and evaluating the performance of spin- and valley-dependent devices. It provides direct information on the ease with which charge carriers propagate through the system under the influence of external conditions. In Landauer's formalism \cite{Landauer1957,Langreth1981}, conductance is determined by the transmission properties of the system and is obtained by integrating the transmission probability over all available transport channels. Taking into account the spin and valley degrees of freedom, the total conductance can be expressed as
\begin{align}\label{Con}
G = \frac{e^2}{h} \sum_{\tau,s_z}
\int T_{\tau s_z}(E,k_y, V_0, L, \varepsilon)\, dk_y,
\end{align}
where $T_{\tau s_z}$ is the transmission probability \eqref{TTPP} for electrons with valley index $\tau$ and spin index $s_z$, and  ${e^2}/{h}$ is the conductance quantum, which sets the fundamental scale for electronic transport. 


To quantify valley selectivity, we define valley polarization as the relative contrast between the conductances associated with the $K$ and $K'$ valleys. It lets us see how much the system prefers transport in a specific valley. It is written as
	\begin{align}\label{Pv}
		P_v = \frac{G_K - G_{K'}}{G_K + G_{K'}}.
	\end{align}
Similarly, spin polarization is introduced to characterize the dependence of electronic transport on the spin orientation. It measures the relative difference between the contributions of spin-up and spin-down electrons. This is given by
\begin{align}\label{Ps}
P_s = \frac{G_\uparrow - G_\downarrow}{G_\uparrow + G_\downarrow}.
\end{align}
As a result, we can have the conductances for the two spin states and the two valleys, respectively,
\begin{align}\label{Csv}
&G_{\uparrow(\downarrow)}= \frac{G_{K\uparrow(\downarrow)} + G_{K'\uparrow(\downarrow)}}{2},\\
&G_{K(K')} = \frac{G_{K(K')\uparrow} + G_{K(K')\downarrow}}{2}.
\end{align}
We present the numerical results for conductance, spin polarization, and valley polarization below. We then analyze how these results are affected by strain, barrier height and width, and incident energy of charge carriers.

\begin{figure}[ht]
\subfloat[]{\includegraphics[scale=0.28]{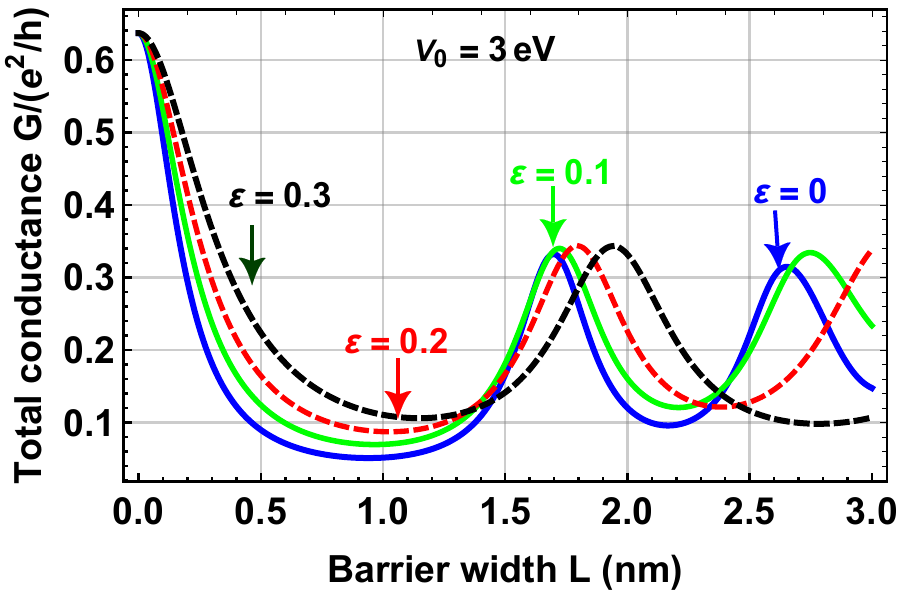}\label{f6a}}
\subfloat[]{\includegraphics[scale=0.28]{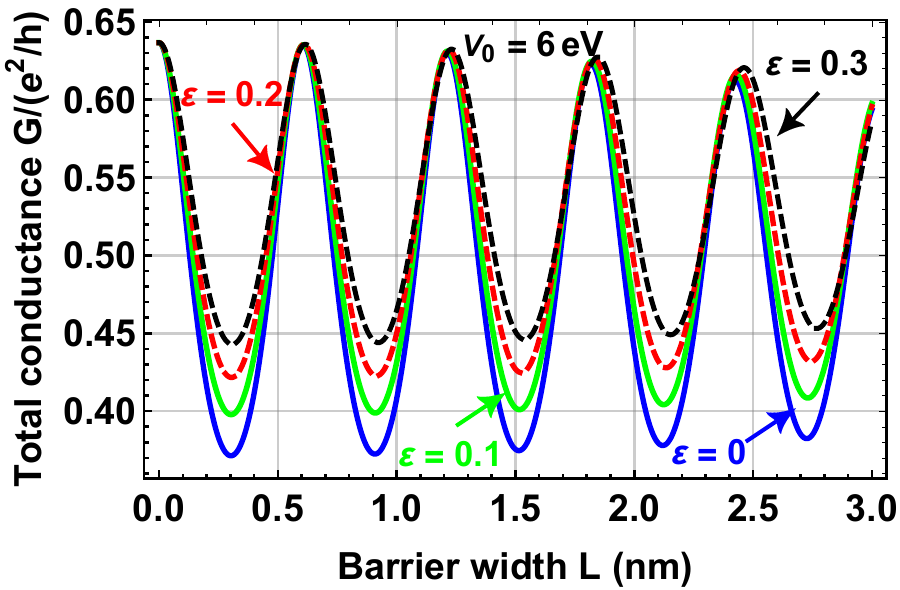}\label{f6b}}
\caption{Total conductance  versus the barrier width $L$ for $E=4$ eV, $\varepsilon=(0,0.1,0.2,0.3)$, and two barrier height values  \text{{(a)}}: $V_0=3$ \text{eV}, \text{{(b)}}: $V_0=6$ \text{eV}.}\label{z4}
\end{figure}

In Fig.~\ref{z4}, we present the total conductance versus the barrier width 
$L$ for $E = 4$~eV and four distinct strain values $\varepsilon = 0, 0.1, 
0.2, 0.3$. We choose $V_0 = 3$~eV in Fig.~\ref{f6a} and $V_0 = 6$~eV in 
Fig.~\ref{f6b}. We observe that in the case without deformation 
($\varepsilon = 0$), the total conductance exhibits quasi-periodic oscillations 
characteristic of Fabry--P\'erot interference, resulting from multiple 
reflections of carriers inside the barrier --- a phenomenon widely reported 
in Dirac-type systems \cite{Young2009,Krstajic2016}. {These 
	oscillations arise because the conductance, obtained by integrating the 
	transmission probability over all transverse channels $k_y$ via 
	Eq.~(\ref{Con}), inherits the resonance structure of $T_{\tau s_z}$: 
	constructive interference occurs whenever the accumulated phase $q_x L = 
	n\pi$ is satisfied, producing conductance maxima at discrete values of $L$ 
	separated by $\Delta L = \pi / q_x$.} These oscillations are modified in 
amplitude and phase when deformation is introduced ($\varepsilon \neq 0$). 
{Physically, the strain-induced gauge field $A_x = 
	\beta\varepsilon(1+\nu)$ modifies the longitudinal wave vector inside the 
	barrier as $q_x \rightarrow q_x + \tau A_x$, so that the resonance condition 
	$q_x L = n\pi$ is now satisfied at a different barrier width $L$, shifting 
	the conductance peaks.} In fact, for $V_0 = 3$~eV in Fig.~\ref{f6a}, the 
increase in $\varepsilon$ leads to a reduction in the conductance and a shift 
in the resonance peaks towards higher values of $L${: since 
	$A_x > 0$ increases $q_x$ in the $K$ valley and decreases it in $K'$, the 
	net $k_y$-integrated conductance experiences a shift in its effective wave 
	vector, requiring a larger $L$ to reach the next resonance condition, 
	consistent with the observed peak displacement. The overall reduction in 
	conductance amplitude reflects the partial cancellation between the 
	$K$ and $K'$ contributions, which oscillate out of phase under strain.} 
For a higher barrier ($V_0 = 6$~eV) in Fig.~\ref{f6b}, the conductance 
remains oscillatory, but the deformation causes a pronounced phase shift of 
the oscillations{: at higher $V_0$, the wave vector mismatch 
	between the incident and barrier regions is larger, making $q_x$ more 
	sensitive to the strain-induced correction $\tau A_x$. A small increment in 
	$\varepsilon$ therefore produces a disproportionately large shift in the 
	accumulated phase $q_x L$, explaining why the phase displacement of the 
	conductance oscillations is more pronounced at $V_0 = 6$~eV than at 
	$V_0 = 3$~eV.} This indicates that the resonant states are highly sensitive 
to mechanical strain. Compared to the case without deformation 
($\varepsilon = 0$), this behavior reveals a break in propagation symmetry 
and a change in the quantum resonance conditions across the barrier. 
{This symmetry breaking is a direct consequence of the 
	time-reversal-preserving but valley-antisymmetric nature of the gauge field 
	$A_x$: since $A_x$ enters with opposite signs for $K$ and $K'$, it lifts 
	the valley degeneracy of the conductance without violating time-reversal 
	symmetry, which would require $G_K(A_x) = G_{K'}(-A_x)$.} An increase in 
deformation and barrier potential simultaneously modifies electronic states 
and accumulated phases, impacting conductance and quantum interference. 
{Specifically, $\varepsilon$ controls the position of the 
	conductance peaks in $L$-space through the phase $\tau A_x L$, while $V_0$ 
	controls their amplitude through the transmission prefactor 
	$\cos\theta\cos\phi$ in Eq.~(\ref{eq35}), confirming that strain and gating 
	act as orthogonal and independently operable tuning parameters for 
	conductance engineering in monolayer MoS$_2$.}

\begin{figure}[ht]
\subfloat[]{\includegraphics[scale=0.28]{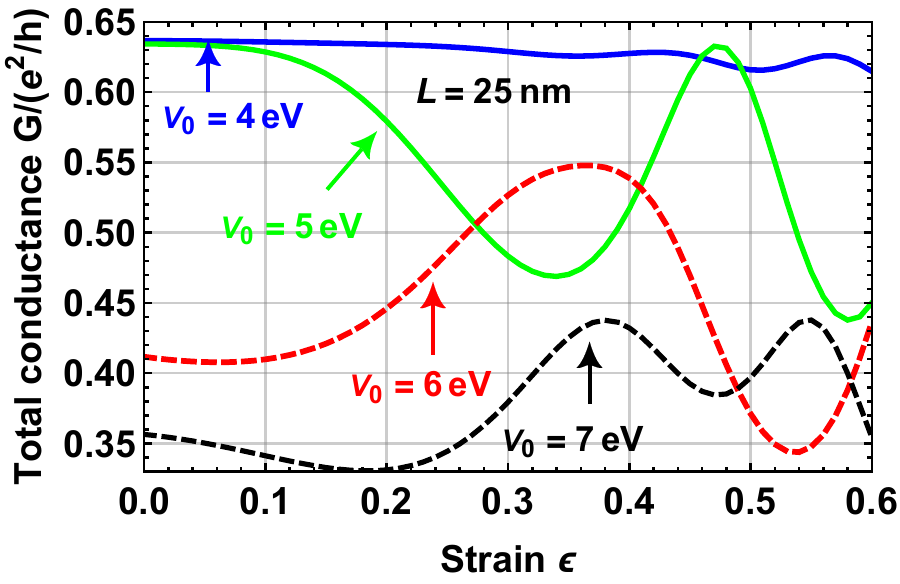}\label{f7a}}
\subfloat[]{\includegraphics[scale=0.28]{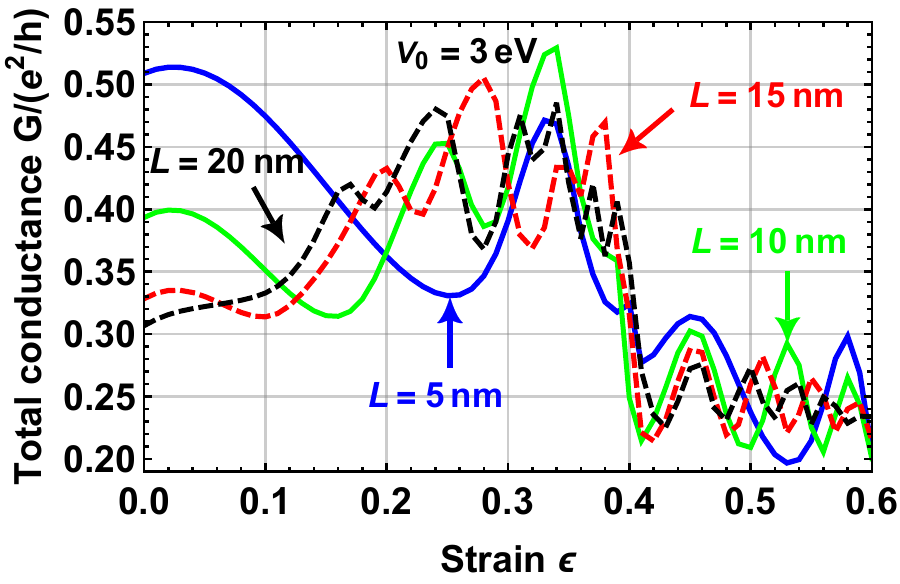}\label{f7b}}
\caption{Total conductance  versus the strain $\varepsilon$ for $E=2$ eV.   \text{{(a)}}: $L=25$ \text{nm},  $V_0=(4, 5, 6, 7)$ \text{eV}, and \text{{(b)}}: $V_0=3$ \text{eV},  $L=(5, 10, 15, 20)$ \text{nm}.}\label{zr}
\end{figure}

Figure~\ref{zr} shows the total conductance versus the strain $\varepsilon$ 
for energy $E = 2$~eV. In Fig.~\ref{f7a}, the barrier length is fixed at 
$L = 25$~nm, while the barrier height varies $V_0 = (4, 5, 6, 7)$~eV. We 
observe that conductance decreases overall as $V_0$ increases, which is 
expected since a higher barrier reduces electron transmission \cite{Cheng2015}. 
{This overall reduction follows directly from the transmission 
	formula Eq.~(\ref{eq35}): raising $V_0$ increases the wave vector mismatch 
	between the incident region ($k_x$) and the barrier region ($q_x$), which 
	reduces the angles $\theta$ and $\phi$ satisfying the resonance condition and 
	suppresses the prefactor $\cos\theta\cos\phi$, leading to a net decrease in 
	the $k_y$-integrated conductance.} However, the conductance does not decrease 
monotonically with deformation. It exhibits marked oscillations due to 
Fabry--P\'erot-type quantum interference within the barrier. 
{These oscillations arise because the strain-induced gauge 
	field $A_x = \beta\varepsilon(1+\nu)$ continuously shifts the longitudinal 
	wave vector inside the barrier as $q_x \rightarrow q_x + \tau A_x$, cycling 
	the system through successive constructive and destructive interference 
	conditions $q_x L = n\pi$ as $\varepsilon$ increases. Each time the phase 
	condition is met for a given spin--valley channel $(\tau, s_z)$, a 
	conductance maximum appears, while destructive interference between channels 
	produces the minima.} The deformation alters the energy spectrum and the 
longitudinal wave vector, which shifts the resonance conditions and 
alternately causes maxima and minima in conductance. {Since 
	the four spin--valley channels $(\tau = \pm 1, s_z = \pm 1)$ experience 
	different gauge shifts $\tau A_x$ and different spin--orbit energies 
	$\Delta_{\tau s_z} = \Delta - 2\lambda\tau s_z$, their resonance conditions 
	are met at distinct values of $\varepsilon$, producing a superposition of 
	oscillations with incommensurate periods that gives rise to the complex, 
	apparently irregular conductance pattern observed at large $\varepsilon$.} 
In Fig.~\ref{f7b}, the barrier is maintained at a height $V_0 = 3$~eV while 
its width takes the values $L = (5, 10, 15, 20)$~nm. One can see that the 
number of conductance oscillations clearly increases with the width of the 
barrier $L$, as confirmed in \cite{Krstajic2016,ElAitouni2025}. 
{This can be understood quantitatively from the resonance 
	condition $q_x(\varepsilon) L = n\pi$: for a fixed increment $\Delta\varepsilon$ 
	in strain, the phase advance is $\Delta\phi = (\partial q_x/\partial A_x) 
	\cdot (\partial A_x/\partial\varepsilon) \cdot \Delta\varepsilon \cdot L = 
	\beta(1+\nu) L \cdot \Delta\varepsilon$, which scales linearly with $L$. 
	A longer barrier therefore accumulates phase more rapidly with strain, 
	fitting more complete oscillation cycles within the same range of 
	$\varepsilon$ and producing the higher oscillation density observed for 
	$L = 20$~nm relative to $L = 5$~nm.} This increase can be explained by 
the accumulation of a larger phase of the electronic states as they cross 
a longer barrier, which complicates the conditions for quantum resonance. 
Furthermore, the increase in deformation $\varepsilon$ modifies the amplitude 
and phase shift of the oscillations, displacing the resonance peaks towards 
different values of $\varepsilon${: while $L$ controls the 
	oscillation frequency through the phase accumulation rate $\beta(1+\nu)L$, 
	strain $\varepsilon$ acts as the swept variable that drives the system through 
	these resonances, and the barrier height $V_0$ sets the overall conductance 
	envelope by controlling the transmission amplitude at each resonance. These 
	three parameters therefore play distinct and non-redundant roles in shaping 
	the conductance landscape, confirming the orthogonal tunability of the 
	system.}

\begin{figure}[ht]
\subfloat[]{\includegraphics[scale=0.28]{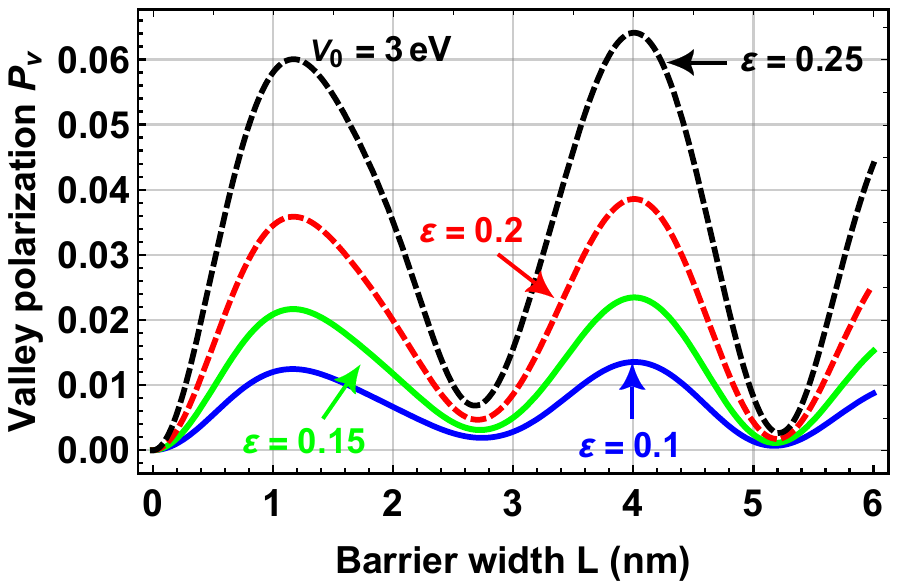}\label{f8a}}
\subfloat[]{\includegraphics[scale=0.28]{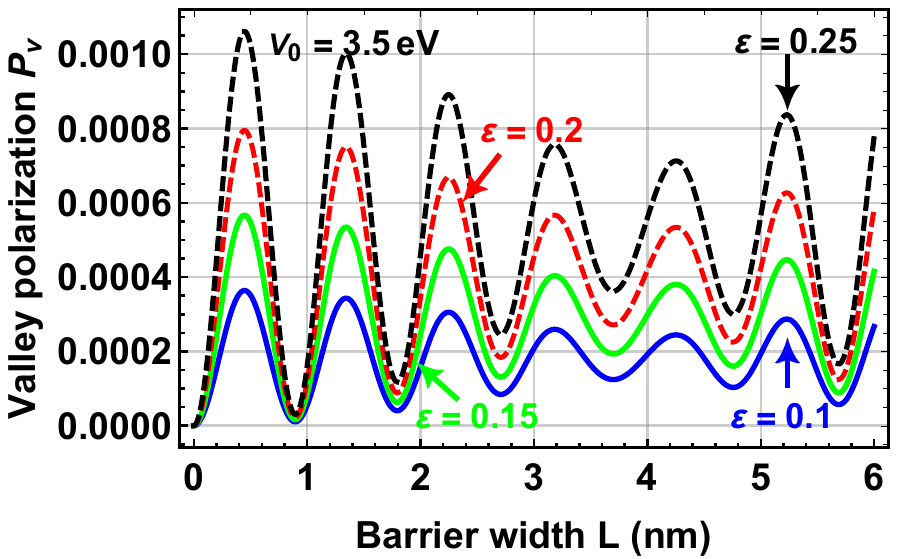}\label{f8b}}
\caption{Valley polarization $P_v$ versus the barrier width $L$ for $E=2$ eV, $\varepsilon=(0.1, 0.15, 0.2, 0 .25)$,  and two barrier height values   \text{{(a)}}: $V_0=3$ \text{eV}, \text{{(b)}}: $V_0=3.5$ \text{eV}.}\label{znA}
\end{figure}

Figure~\ref{znA} presents the valley polarization $P_v$ versus the barrier 
width $L$ for $E = 2$~eV, different strain values $\varepsilon = (0.1, 0.15, 
0.2, 0.25)$, and two barrier heights: $V_0 = 3$~eV in Fig.~\ref{f8a} and 
$V_0 = 3.5$~eV in Fig.~\ref{f8b}. Indeed, we notice that $P_v$ varies 
oscillatorily with $L${: since $P_v = (G_K - G_{K'}) / 
	(G_K + G_{K'})$, any difference in the Fabry--P\'erot resonance conditions 
	between the two valleys directly translates into a nonzero and oscillatory 
	$P_v$. The $K$ and $K'$ conductances oscillate with the same frequency but 
	with a phase difference $\Delta\phi = 2A_x L$ imposed by the valley-dependent 
	gauge field, so that $P_v$ oscillates at twice the frequency of the individual 
	conductance oscillations, with an amplitude set by the degree of phase 
	mismatch between the two valleys.} This behavior is attributed to Fabry--P\'erot 
quantum interference arising from multiple carrier reflections inside the 
barrier, as found in \cite{Krstajic2016,Surrente2017}. The amplitude of the 
oscillations clearly increases with $\varepsilon${: a larger 
	strain increases $A_x = \beta\varepsilon(1+\nu)$, which widens the phase 
	difference $\Delta\phi = 2A_x L$ between the $K$ and $K'$ resonance 
	conditions, lifting their degeneracy more strongly and producing a larger 
	contrast between $G_K$ and $G_{K'}$. In the limit $\varepsilon \rightarrow 0$, 
	$A_x \rightarrow 0$ and the two valleys become degenerate, so that $P_v 
	\rightarrow 0$ identically, consistent with the near-zero polarization observed 
	at small strain values.} Moreover, increasing $L$ leads to an increase in the 
number of oscillations and a shift in the polarization peaks{: 
	from the resonance condition $q_x L = n\pi$, a longer barrier requires a 
	smaller increment in $L$ to advance from one resonance to the next, 
	compressing the oscillation period in $L$-space and increasing the fringe 
	density. Simultaneously, the phase mismatch $\Delta\phi = 2A_x L$ grows 
	linearly with $L$, shifting the polarization peaks to smaller values of $L$ 
	as $\varepsilon$ increases, in agreement with the leftward displacement of 
	peaks observed with increasing strain.} This demonstrates that $L$ controls 
the phase accumulated by the wavefunctions and the resonance conditions. 
Finally, a comparison between Figs.~\ref{f8a} and~\ref{f8b} reveals the 
impact of the barrier height $V_0$. A higher barrier modifies the structure 
and amplitude of the oscillations{: raising $V_0$ from 
	3~eV to 3.5~eV increases the wave vector mismatch between the incident and 
	barrier regions, which reduces the transmission amplitude $\cos\theta\cos\phi$ 
	in Eq.~(\ref{eq35}) and suppresses both $G_K$ and $G_{K'}$. However, since 
	the suppression is not identical for the two valleys --- the strain-shifted 
	wave vectors $q_x + A_x$ and $q_x - A_x$ respond differently to the increased 
	confinement --- the relative contrast $(G_K - G_{K'})/(G_K + G_{K'})$ is 
	altered, producing the modified oscillation amplitude and structure observed 
	in Fig.~\ref{f8b}. The dramatic reduction in $P_v$ amplitude by approximately 
	two orders of magnitude between $V_0 = 3$~eV and $V_0 = 3.5$~eV signals 
	that the system operates near a confinement threshold at which the two valleys 
	are nearly equally suppressed, making $P_v$ highly sensitive to small changes 
	in barrier height in this parameter regime.}

\begin{figure}[ht]
\subfloat[]{\includegraphics[scale=0.28]{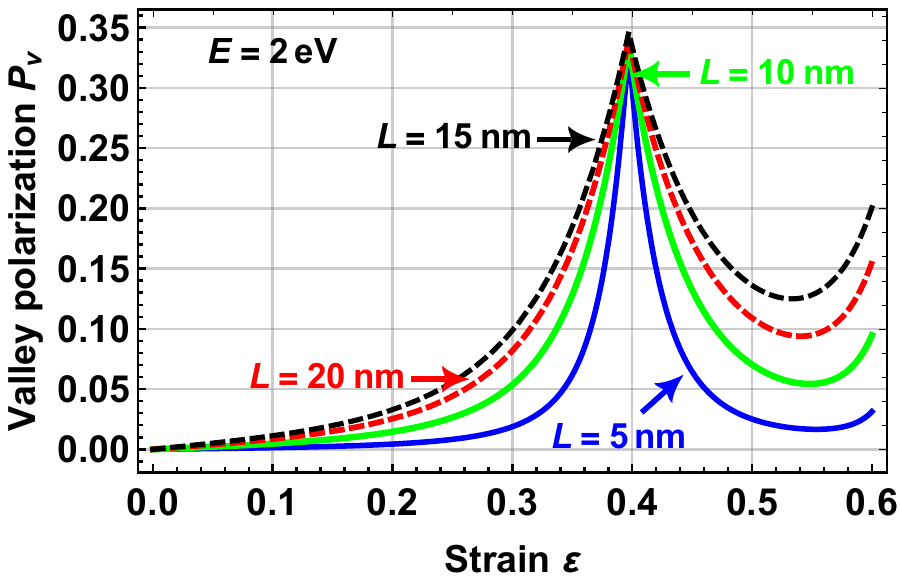}\label{f9a}}
\subfloat[]{\includegraphics[scale=0.28]{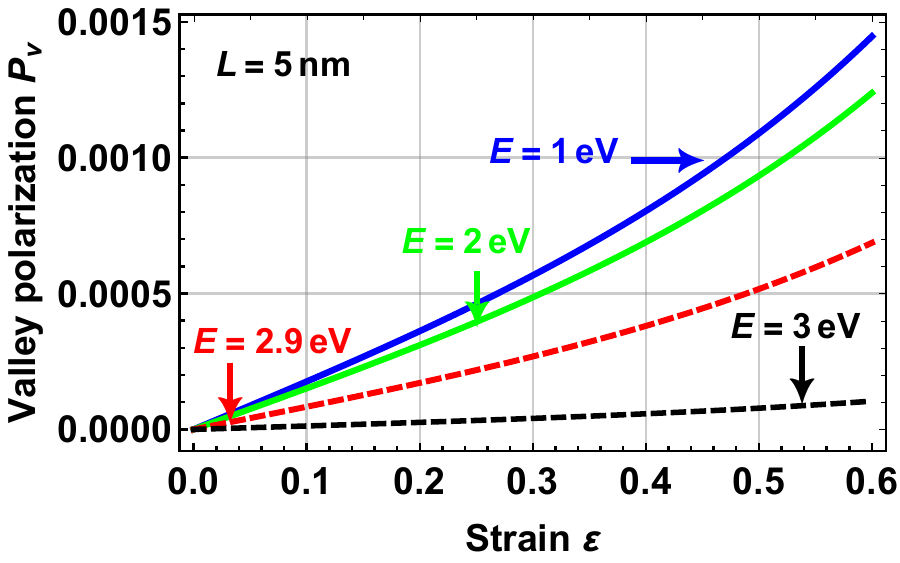}\label{f9b}}
\caption{Valley polarization $P_v$ versus the strain $\varepsilon$ for $V_0=3$ eV.  \text{{(a)}}: $E=2$ eV, $L=(5,10, 15, 20)$ \text{nm},  \text{{(b)}}: $L=5$ \text{nm}, $E=(1, 2, 2.9, 3)$ {eV}.} \label{zv}
\end{figure}

Figure~\ref{zv} illustrates the evolution of the valley polarization $P_v$ 
versus the strain $\varepsilon$ for a barrier height $V_0 = 3$~eV. In 
Fig.~\ref{f9a} for $E = 2$~eV, $L = (5, 10, 15, 20)$~nm, we observe that 
$P_v$ is almost zero in the absence of deformation, reflecting the equivalence 
of the $K$ and $K'$ valleys{: at $\varepsilon = 0$, the gauge 
	field $A_x = \beta\varepsilon(1+\nu) = 0$ vanishes identically, so that the 
	Hamiltonian recovers its valley degeneracy and the transmission probabilities 
	satisfy $T_{+1,s_z} = T_{-1,s_z}$ for all channels, yielding $G_K = G_{K'}$ 
	and hence $P_v = 0$ exactly.} In contrast, once deformation is applied, $P_v$ 
increases rapidly and attains a maximum for an intermediate value of 
$\varepsilon${: as $\varepsilon$ grows from zero, $A_x$ 
	increases and the valley-dependent phase shift $\Delta\phi = 2A_x L$ lifts 
	the degeneracy between $G_K$ and $G_{K'}$, driving $P_v$ away from zero. 
	The subsequent decline of $P_v$ beyond the maximum arises because, at large 
	$\varepsilon$, the gauge field shifts both valleys far from their respective 
	resonance conditions, reducing the absolute transmission in both valleys and 
	causing their conductances to converge again, thereby suppressing $P_v$.} 
The maximum amplitude of polarization increases with the barrier width $L$ 
\cite{Surrente2017}{: from the phase mismatch $\Delta\phi = 
	2A_x L$, a longer barrier amplifies the valley-dependent phase difference for 
	the same strain value, producing a larger contrast between $G_K$ and $G_{K'}$ 
	and hence a higher peak $P_v$. Furthermore, since the resonance condition 
	$q_x L = n\pi$ is satisfied at smaller $\varepsilon$ for larger $L$, the 
	polarization peak shifts to lower strain values as $L$ increases, consistent 
	with the leftward displacement of the maxima observed in Fig.~\ref{f9a}.} 
This indicates that quantum interference effects within the deformed region 
play a crucial role in valley filtering. By fixing $L = 5$~nm and varying 
$E = (1, 2, 2.9, 3)$~eV as shown in Fig.~\ref{f9b}, one can clearly see 
that the valley polarization is greater at low energy and gradually decreases 
as the energy increases, in agreement with literature \cite{Kioseoglou2012}. 
{This energy dependence can be understood from the 
	transmission formula Eq.~(\ref{eq35}): at low $E$, the incident wave vector 
	$k_x$ is small and the barrier represents a strong perturbation, so that 
	the valley-dependent gauge shift $\tau A_x$ produces a large relative change 
	in the transmission probability between $K$ and $K'$, yielding high $P_v$. 
	As $E$ increases, $k_x$ grows and the relative importance of $A_x$ as a 
	fraction of $k_x$ diminishes, reducing the transmission asymmetry between 
	the two valleys. In the limit $E \gg V_0$, the barrier becomes effectively 
	transparent for all channels regardless of valley index, so that 
	$T_{+1,s_z} \approx T_{-1,s_z} \approx 1$ and $P_v \rightarrow 0$.} At 
high energy, $P_v$ becomes very small, even for high values of deformation, 
which can be explained by the fact that carriers cross the barrier with an 
almost unitary probability{, confirming that low incident 
	energy is the necessary operating condition for efficient valley filtering 
	in this system, and that the valley filter can be switched off simply by 
	raising the carrier energy through electrostatic gating.}


\begin{figure}[ht]
\subfloat[]{\includegraphics[scale=0.28]{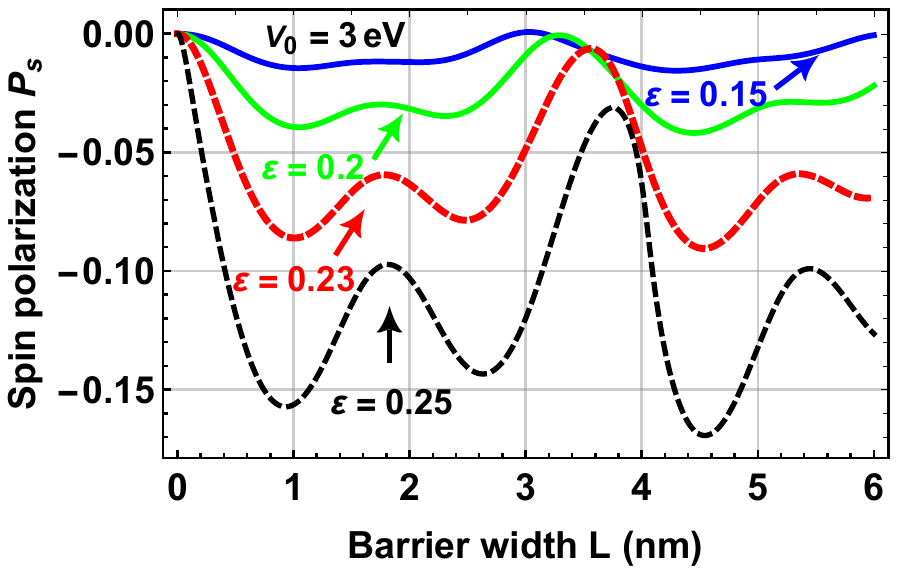}\label{f10a}}\subfloat[]{\includegraphics[scale=0.28]{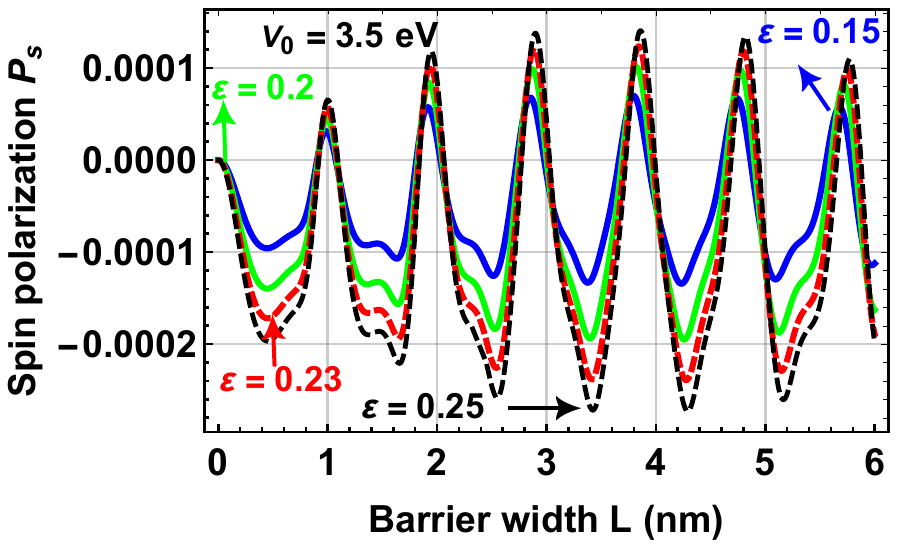}\label{f10b}}
\caption{Spin polarization $P_s$ versus the barrier width $L$ for $E=2$ eV, $\varepsilon=(0.15, 0.2, 0.23, 0 .25)$  and two barrier height values   \text{{(a)}}: $V_0=3$ \text{eV}, \text{{(b)}}: $V_0=3.5$ \text{eV}.} \label{zn}
\end{figure}

To investigate the influence of strain and the potential barrier height on 
spin polarization $P_s$, in Fig.~\ref{zn} we show $P_s$ versus the barrier 
width $L$ for different values of $\varepsilon = (0.15, 0.2, 0.23, 0.25)$, 
$E = 2$~eV, and for two barrier heights: $V_0 = 3$~eV in Fig.~\ref{f10a} 
and $V_0 = 3.5$~eV in Fig.~\ref{f10b}. In both cases, we observe oscillatory 
behavior in $P_s$ with respect to $L$, which is characteristic of 
Fabry--P\'erot-type quantum interference of electronic states confined within 
the barrier region. This phenomenon has been widely discussed in Dirac-type 
systems \cite{Krstajic2016,Tworzydlo2006,Young2009,Rickhaus2013}. 
{The origin of these oscillations in $P_s$ lies in the 
	spin-dependent phase accumulation inside the barrier: the spin--orbit coupling 
	term $-\lambda\tau s_z$ assigns different effective band gap parameters 
	$\Delta_{\tau s_z} = \Delta - 2\lambda\tau s_z$ to the spin-up and spin-down 
	channels, so that their longitudinal wave vectors $q_x^{(\uparrow)}$ and 
	$q_x^{(\downarrow)}$ differ inside the barrier. As $L$ increases, the two 
	spin channels accumulate distinct phases $q_x^{(\uparrow)}L$ and 
	$q_x^{(\downarrow)}L$, satisfying the Fabry--P\'erot condition $q_x L = 
	n\pi$ at different values of $L$ and producing the oscillatory behavior of 
	$P_s = (G_\uparrow - G_\downarrow)/(G_\uparrow + G_\downarrow)$ observed 
	in both figures.} An increase in strain improves the amplitude of $P_s$, 
resulting in more pronounced negative $P_s$ values. {The 
	consistently negative sign of $P_s$ reflects the fact that, for the 
	parameter range considered, the spin-down channel ($s_z = -1$) lies closer 
	to its Fabry--P\'erot resonance condition than the spin-up channel, yielding 
	$G_\downarrow > G_\uparrow$ and hence $P_s < 0$ throughout. As $\varepsilon$ 
	increases, the gauge field $A_x = \beta\varepsilon(1+\nu)$ enters the 
	longitudinal wave vector as $q_x^{(s_z)} = \tau\sqrt{\kappa_{\tau s_z}}$, 
	where $\kappa_{\tau s_z}$ depends on both $\Delta_{\tau s_z}$ and $A_x$. 
	The combined effect of spin--orbit splitting and strain-induced gauge shift 
	causes the wave vectors $q_x^{(\uparrow)}$ and $q_x^{(\downarrow)}$ to 
	diverge more strongly with increasing $\varepsilon$, amplifying the phase 
	mismatch between the two spin channels and deepening the negative trough 
	in $P_s$.} This indicates stronger separation of the spin channels, which 
is explained by the interaction between the strain and the strong intrinsic 
spin--orbit coupling of MoS$_2$. The resulting different propagation phases 
for electrons with opposite spins break the spin-transport symmetry
{: this symmetry breaking is protected by the interplay 
	between time-reversal symmetry and the valley degree of freedom. At 
	$\varepsilon = 0$, the contributions from the $K$ and $K'$ valleys to 
	$G_\uparrow$ and $G_\downarrow$ partially cancel due to the spin--valley 
	locking relation $\Delta_{\tau s_z} = \Delta_{-\tau,-s_z}$, keeping $P_s$ 
	small. The application of strain breaks this cancellation by introducing 
	the valley-antisymmetric shift $\tau A_x$, which lifts the spin--valley 
	locking compensation and allows $P_s$ to develop a finite and growing 
	amplitude.} Furthermore, comparing $V_0 = 3$~eV and $V_0 = 3.5$~eV shows 
that increasing the barrier height reduces the average amplitude of the spin 
polarization while increasing the oscillation frequency{: 
	the frequency increase follows from the compression of the resonance spacing 
	$\Delta L = \pi / q_x$ at higher $V_0$, where the larger wave vector mismatch 
	between incident and barrier regions increases $q_x$ inside the barrier and 
	reduces the period of the Fabry--P\'erot fringes. The amplitude reduction, 
	on the other hand, reflects the more uniform suppression of both spin 
	conductances under stronger confinement: at $V_0 = 3.5$~eV, the barrier 
	attenuates $G_\uparrow$ and $G_\downarrow$ more equally, reducing their 
	relative difference and hence $|P_s|$. The reduction in $|P_s|$ by 
	approximately three orders of magnitude between $V_0 = 3$~eV and 
	$V_0 = 3.5$~eV signals that the system is close to a spin-degenerate 
	confinement regime at the higher barrier, where spin selectivity is 
	strongly suppressed and the device transitions from a spin-active to a 
	spin-neutral transport regime.} This is attributed to stronger confinement 
and a modification of the resonance conditions inside the barrier 
\cite{Schaibley2016}.

\begin{figure}[ht]
\subfloat[]{\includegraphics[scale=0.28]{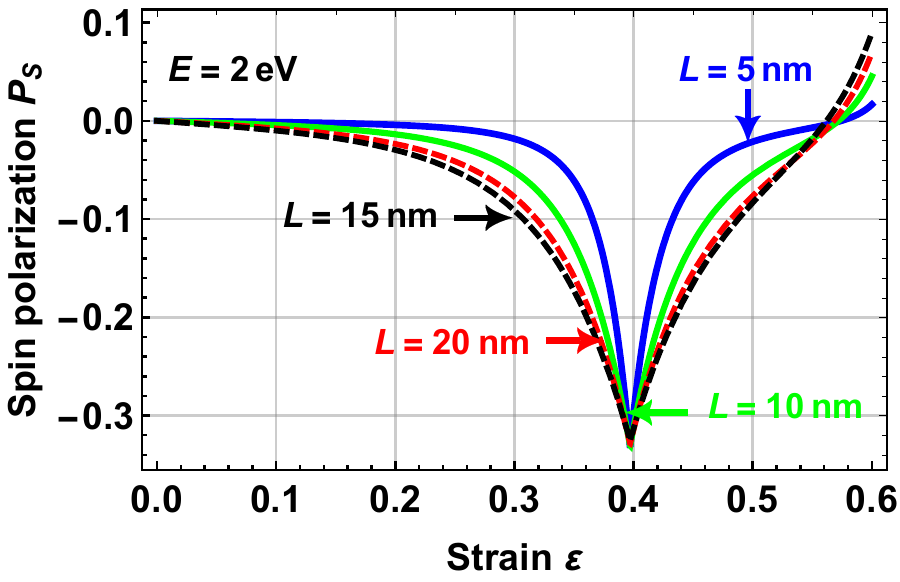}\label{f11a}}
\subfloat[]{\includegraphics[scale=0.28]{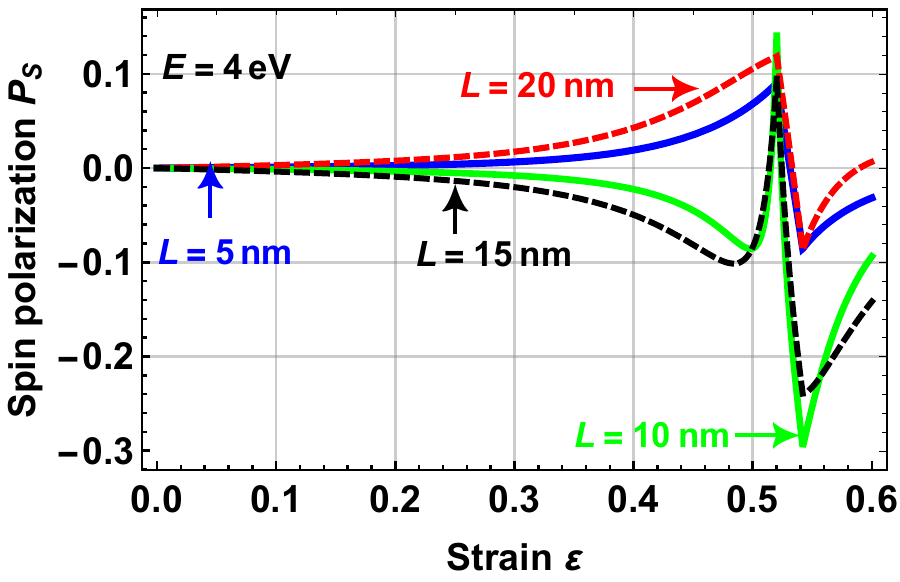}\label{f11b}}
\caption{Spin polarization $P_s$ versus the strain $\varepsilon$ for $V_0=3$ eV, $L=5$ \text{nm}, $10$ \text{nm}, $15$ \text{nm}, $20$ \text{nm} and two energy values \text{{(a)}}: $E=2$ \text{eV}, \text{{(b)}}: $E=4$ \text{eV}. } \label{zk}
\end{figure}

Figure~\ref{zk} depicts the spin polarization $P_s$ versus the strain 
$\varepsilon$ for $V_0 = 3$~eV, four barrier widths $L = (5, 10, 15, 20)$~nm 
and two distinct incident energies: $E = 2$~eV in Fig.~\ref{f11a} and 
$E = 4$~eV in Fig.~\ref{f11b}. In both cases, we observe that $P_s$ is zero 
in the absence of deformation, reflecting the initial degeneracy of the spin 
channels: at $\varepsilon = 0$, the gauge field $A_x = \beta\varepsilon(1+\nu)$ 
vanishes and the Hamiltonian recovers its full spin--valley symmetry. The 
spin--valley locking relation $\Delta_{\tau s_z} = \Delta_{-\tau,-s_z}$ then 
ensures that the contribution of channel $(+1,\uparrow)$ to the conductance 
is exactly compensated by channel $(-1,\downarrow)$, and similarly for 
$(+1,\downarrow)$ and $(-1,\uparrow)$, so that $G_\uparrow = G_\downarrow$ 
and $P_s = 0$ exactly. As the strain increases, $P_s$ becomes nonzero and 
exhibits a strongly nonlinear behavior. This agrees with the intrinsic symmetry 
expected in an unstrained system, as discussed in studies on the spin--valley 
physics of MoS$_2$ \cite{int3,Zhu2011}. The nonlinearity of $P_s$ as a 
function of $\varepsilon$ reflects the fact that the gauge field $A_x \propto 
\varepsilon$ enters the transmission probability through the transcendental 
resonance condition $\cos(q_x L)\cos\theta\cos\phi - i\sin(q_x L)(1 - 
\sin\theta\sin\phi) = 0$, where $q_x$ depends nonlinearly on $A_x$ through 
Eq.~(\ref{eq12}), so that $P_s(\varepsilon)$ is a nonlinear and non-monotonic 
function of strain even though $A_x$ itself is linear in $\varepsilon$. 
{It is worth noting that this nonlinearity becomes more 
	pronounced at larger $L$, since the phase $q_x L$ is more sensitive to 
	changes in $q_x$ for longer barriers, amplifying the effect of any 
	strain-induced modification of the wave vector.} More precisely, 
Fig.~\ref{f11a} shows that, for low $E$, $P_s$ becomes negative as 
$\varepsilon$ increases and reaches a pronounced minimum around a critical 
strain value, before rising rapidly for higher strains. The critical strain 
$\varepsilon^*$ at which the minimum occurs corresponds to the value of $A_x$ 
for which the phase mismatch between spin-up and spin-down channels, $\delta\phi 
= [q_x^{(\uparrow)} - q_x^{(\downarrow)}]L$, reaches $\pi/2$: at this point 
the destructive interference between the two spin channels is maximized, 
producing the deepest negative $P_s$. Beyond $\varepsilon^*$, the phase 
mismatch exceeds $\pi/2$ and the interference becomes constructive again, 
causing $P_s$ to rise back toward zero. {This behavior is 
	reminiscent of a spin-dependent Fabry--P\'erot resonator in which the strain 
	plays the role of a phase knob: a quarter-period advance of the relative 
	phase $\delta\phi$ between spin channels drives the system from zero to 
	maximum spin polarization, while a half-period advance returns it to zero 
	with opposite sign, in direct analogy with a two-beam interference pattern.} 
This minimum is more pronounced for larger barrier widths, especially for 
$L = 10$~nm and $L = 20$~nm: since the phase mismatch $\delta\phi = 
[q_x^{(\uparrow)} - q_x^{(\downarrow)}]L$ scales linearly with $L$, a longer 
barrier amplifies the spin-channel phase difference for the same $\varepsilon$, 
producing a deeper and sharper minimum in $P_s$ at a smaller critical strain 
$\varepsilon^*$. This explains both the deepening of the minimum and its 
leftward shift with increasing $L$ observed in Fig.~\ref{f11a}. Physically, 
this effect originates from spin-dependent resonances in electron transmission, 
resulting from the accumulation of different phases for electrons with opposite 
spins as they propagate through the barrier: the spin-up and spin-down channels 
acquire phases $q_x^{(\uparrow)}L$ and $q_x^{(\downarrow)}L$ respectively, 
where the difference $q_x^{(\uparrow)} - q_x^{(\downarrow)}$ is controlled 
by the combined action of the spin--orbit splitting $2\lambda\tau$ and the 
strain-induced gauge field $\tau A_x$, both of which are valley-dependent and 
add constructively in one valley while partially canceling in the other. 
{Summing over both valleys in the conductance integral 
	Eq.~(\ref{Con}) therefore produces a net spin polarization that reflects 
	the incomplete cancellation between the $K$ and $K'$ contributions, with 
	the degree of cancellation controlled by $\varepsilon$ through $A_x$.}
In contrast, Fig.~\ref{f11b}, obtained for a higher $E$, shows qualitatively 
different behavior. For small $\varepsilon$, the spin polarization remains 
close to zero, particularly at low values of $L$: at high energy $E = 4$~eV, 
the incident wave vector $k_x$ is large and the transmission probability is 
close to unity for all spin--valley channels, so that $G_\uparrow \approx 
G_\downarrow$ and $P_s \approx 0$ regardless of strain, until $A_x$ becomes 
large enough to shift one spin channel into a new resonance condition while 
the other remains off-resonance. {The threshold strain above 
	which $P_s$ departs significantly from zero therefore increases with $E$, 
	since a larger $k_x$ requires a proportionally larger $A_x$ --- and hence 
	a larger $\varepsilon$ --- to produce a significant relative shift between 
	the spin channels. This explains why the onset of nonzero $P_s$ is delayed 
	to higher $\varepsilon$ in Fig.~\ref{f11b} compared to Fig.~\ref{f11a}.} 
However, when $\varepsilon$ exceeds a certain critical value, $P_s$ exhibits 
abrupt variations and sign reversals. These sign reversals signal a qualitative 
change in the transport regime: as $\varepsilon$ crosses a threshold, the 
gauge shift $\tau A_x$ pushes a previously off-resonance spin channel into 
the Fabry--P\'erot condition $q_x L = n\pi$ while simultaneously driving the 
other channel off-resonance, causing $G_\uparrow$ and $G_\downarrow$ to 
exchange dominance and reversing the sign of $P_s$. At high energy, multiple 
such crossings can occur within the strain range considered, because the larger 
$k_x$ means that smaller increments in $A_x$ are sufficient to cycle through 
resonance conditions, producing the rapid and repeated sign reversals observed 
in Fig.~\ref{f11b}. This behavior constitutes the electrostatic spin inversion 
predicted in this work: the sign of the spin polarization --- and hence the 
dominant spin channel of the transmitted current --- can be reversed purely 
by adjusting the strain or the gate voltage, without any geometric modification 
of the device. These rapid changes reflect the emergence of new resonance 
conditions in the barrier, associated with the selective opening and closing 
of transmission channels for each spin. {Taken together, 
	Figs.~\ref{f11a} and~\ref{f11b} demonstrate that incident energy $E$ and 
	barrier width $L$ provide complementary handles on the spin polarization: 
	$L$ controls the sensitivity of $P_s$ to strain through the phase 
	accumulation rate, while $E$ sets the strain threshold above which spin 
	selectivity emerges, offering a two-parameter space for the engineering of 
	spin-polarized currents in monolayer MoS$_2$.}

Although the polarization values are not always close to unity, these finite values indicate a selective contribution of different spin and valley channels to the transmitted current. The variation of the polarization with strain and scalar potential reflects the tunability of the transport properties through external parameters. In realistic devices, such intermediate polarization regimes are important because they allow continuous control of the spin and valley currents rather than only switching between fully polarized and unpolarized states. Therefore, the present results demonstrate that mechanical strain and electrostatic engineering provide flexible tools for manipulating spin- and valley-dependent transport in monolayer MoS$_2$.

\section{Experimental feasibility and limitations}\label{V}

{From an experimental standpoint, the two key ingredients 
	of our model,  a tunable electrostatic barrier and a controllable uniaxial 
	strain field,  have both been demonstrated independently in monolayer 
	MoS$_2$ and related transition metal dichalcogenides. Electrostatic barriers 
	can be realized using local top gates or split-gate geometries, as routinely 
	employed in van der Waals heterostructure devices~\cite{int1,intr1,intr2}. 
	Uniaxial strain can be applied by transferring MoS$_2$ onto flexible 
	substrates such as PDMS or PET and mechanically stretching the 
	substrate~\cite{int17,int18,int19}, or through nanoscale probe techniques 
	such as AFM tips~\cite{int20,int22}. Spin- and valley-resolved transport 
	signals have been detected experimentally through polarization-resolved 
	photoluminescence and magneto-transport 
	measurements~\cite{int3,Zhu2011,Castellanos2013}.}
	
However, several limitations of the present model should be acknowledged. 
	First, we adopt a continuum effective Dirac Hamiltonian, which captures the 
	low-energy physics near the $K$ and $K'$ points accurately but does not 
	account for higher-band corrections or intervalley scattering, which may 
	become relevant at large strain values or high energies. Second, our model 
	assumes a perfectly sharp rectangular barrier profile and a spatially 
	uniform strain field, whereas in practice the electrostatic potential and 
	strain distribution may vary smoothly at the interfaces, which could broaden 
	the Fabry--P\'erot resonances and reduce the sharpness of the spin and 
	valley polarization features. Third, disorder effects such as impurity 
	scattering, substrate-induced potential fluctuations, and grain boundaries 
	are not included in the present treatment; these are known to partially wash 
	out quantum interference effects in real devices~\cite{Young2009,Rickhaus2013}. 
	Fourth, finite-temperature effects are not considered: the sharp Fabry--P\'erot 
	resonances predicted here are most pronounced at low temperature, and thermal 
	broadening of the Fermi distribution is expected to smooth the conductance 
	oscillations and reduce the peak polarization values at room temperature. 
	Despite these idealizations, the qualitative predictions of the model, 
	valley filtering, strain-controlled conductance oscillations, and 
	electrostatic spin inversion, are robust physical effects that survive 
	beyond the strict limits of the model, and the parameter ranges where they 
	are most pronounced ($\varepsilon \leq 0.3$, low incident energy) are 
	accessible with current experimental platforms~\cite{int14,int15,int16,
		int17,int18,int19,Castellanos2013}.

\section{Conclusion}
\label{sec5}

We have theoretically studied electron transport in monolayer MoS$_2$ 
subjected to uniaxial strain in the presence of an electrostatic barrier, 
simultaneously taking into account spin and valley degrees of freedom and 
intrinsic spin--orbit coupling. Using an effective Dirac formalism and 
numerical calculations, we have analyzed the transmission probability, the 
total conductance as well as the spin and valley polarizations as functions 
of various system parameters, including incident energy, angle of incidence, 
barrier height, barrier width, wave vector, and strain. {The 
	central novelty of this work lies in the simultaneous treatment of both 
	perturbations, which reveals physical effects inaccessible when strain or 
	gating are considered independently.}

Our results show that the transmission exhibits pronounced Fabry--P\'erot-type 
resonances, originating from quantum interference and confinement within the 
barrier region. {These resonances arise whenever the 
	accumulated phase satisfies $q_x L = n\pi$, where the longitudinal wave 
	vector $q_x$ depends on the spin--valley channel through $\Delta_{\tau s_z} 
	= \Delta - 2\lambda\tau s_z$, making each of the four spin--valley channels 
	$(\tau = \pm 1, s_z = \pm 1)$ resonate at a distinct set of system 
	parameters.} These resonances are strongly dependent on the spin and valley 
degrees of freedom, and their position and width can be tuned by adjusting 
the strain and the electrostatic parameters.

In the absence of strain, we have found that the transmission probabilities 
are almost identical for the different spin and valley channels, resulting 
in an almost degenerate total conductance. {This degeneracy 
	is protected by the spin--valley locking relation $\Delta_{\tau s_z} = 
	\Delta_{-\tau,-s_z}$, which enforces $G_\uparrow = G_\downarrow$ and 
	$G_K = G_{K'}$ in the unstrained system, yielding $P_s = P_v = 0$ exactly.} 
In contrast, the application of strain profoundly alters the transport 
properties at the transmission level, lifting the degeneracy between the 
$K$ and $K'$ valleys and the spin orientations. {This lifting 
	originates from the valley-dependent gauge field $A_x = \beta\varepsilon
	(1+\nu)$, which shifts the longitudinal wave vector as $q_x \rightarrow 
	q_x + \tau A_x$, breaking the spin--valley locking compensation and driving 
	the four channels into distinct resonance conditions.} This induced asymmetry 
in the transmission is directly reflected in the conductance, which becomes 
spin- and valley-dependent, leading to the emergence of non-zero spin and 
valley polarizations. These polarizations can reach high values and may even 
change sign. {In particular, we have identified two 
	concrete and original predictions of this work: $(i)$ a dual-knob control 
	scheme in which the barrier width $L$ governs the frequency of conductance 
	oscillations through the phase accumulation rate $\beta(1+\nu)L$, while 
	strain independently controls their phase and amplitude through $A_x$; and 
	$(ii)$ electrostatic spin inversion, whereby the sign of $P_s$ --- and hence 
	the dominant spin channel of the transmitted current --- is reversed purely 
	by gate tuning at finite strain, without any geometric reconfiguration of 
	the device. This sign reversal occurs when the gauge shift $\tau A_x$ drives 
	one spin channel into Fabry--P\'erot resonance while simultaneously pushing 
	the other off-resonance, causing $G_\uparrow$ and $G_\downarrow$ to exchange 
	dominance.} Consequently, we have concluded that strain effectively acts as 
a valley-antisymmetric gauge field, separating transport channels according 
to spin and valley, while the electrostatic barrier parameters and the 
incident energy provide additional and orthogonal degrees of freedom to 
finely tune the magnitude and sign of the polarized currents.

Our findings can be realized through experiments on currently available 
platforms, as demonstrated in the studies \cite{int3,Zhu2011,
	Castellanos2013}. The key ingredients in our model, tunable electrostatic 
potentials and controllable strain fields, have been demonstrated in 
atomically thin materials. {The valley polarization is most 
	efficient at low incident energy, where the relative effect of the gauge 
	field $A_x$ on the wave vector is largest, and can be switched off by 
	raising the carrier energy through electrostatic gating. The spin 
	polarization, by contrast, is most sensitive to strain at intermediate 
	barrier widths and exhibits sign reversal at high energy, providing a 
	complementary and independently controllable spin-filtering mechanism.} 
Using these two parameters in one device provides a high level of control 
over the spin and valley degrees of freedom of charge carriers. This makes 
the interplay of strain engineering and potential barriers a promising 
approach for realizing reconfigurable spintronic and valleytronic devices. 
This tunability enables the selective control of spin- and valley-polarized 
currents, paving the way for new possibilities in the development of devices 
in nanotechnology and optoelectronics using 2D materials, where compactness, 
energy efficiency, and functionality are key advantages of this technology.





\end{document}